\documentclass[prx,twocolumn,nofootinbib,citeautoscript,longbibliography,notitlepage,superscriptaddress]{revtex4-1}

\usepackage{graphicx}
\usepackage{dcolumn}
\usepackage{bm}
\usepackage{color}
\usepackage{amsmath}
\usepackage{tabularx,graphicx}
\usepackage{epstopdf}
\usepackage{latexsym}
\usepackage{amssymb}
\usepackage{amsmath}
\usepackage{color, colortbl}
\usepackage{psfrag}
\usepackage{bbm}
\usepackage{bm}
\usepackage{titlesec}
\usepackage{dsfont}
\usepackage{feynmp}
\usepackage{slashed}
\usepackage{multirow}
\usepackage[tight]{subfigure}

\usepackage[papersize={8.5in,11in}]{geometry}

\usepackage{color}

\renewcommand{\epsilon}{\varepsilon}

\allowdisplaybreaks[3]

\begin{document}



\title{Fermion-fermion interaction driven phase transitions in rhombohedral trilayer graphene}

\date{\today}

\author{Qiao-Chu Zhang}
\affiliation{Department of Physics, Tianjin University, Tianjin 300072, P.R. China}

\author{Jing Wang}
\altaffiliation{Corresponding author: jing$\textunderscore$wang@tju.edu.cn}
\affiliation{Department of Physics, Tianjin University, Tianjin 300072, P.R. China}
\affiliation{Tianjin Key Laboratory of Low Dimensional Materials Physics and
Preparing Technology, Tianjin University, Tianjin 300072, P.R. China}

\begin{abstract}
The effects of short-range fermion-fermion interactions on the low-energy properties of rhombohedral trilayer graphene are comprehensively
investigated using the momentum-shell renormalization group method. We take into account all one-loop corrections and establish the
energy-dependent coupled evolutions of independent fermionic couplings that carry the physical information stemming from the interplay of various fermion-fermion interactions. With detailed numerical analysis, we observe that the ferocious competition among all fermion-fermion
interactions can drive fermionic couplings to four distinct fixed points,
dubbed $\textrm{FP}_{1}$, $\textrm{FP}_{2}$, $\textrm{FP}_{3}$, and $\textrm{FP}_{4}$,
in the interaction-parameter space. These fixed points primarily dictate the fate of
the system in the low-energy regime and are always associated with some instabilities characterized
by specific symmetry breakings, leading to certain phase transitions. To determine the favorable states arising from the potential phase transitions,
we introduce a number of fermion-bilinear source terms to characterize the underlying candidate states. By comparing their related susceptibilities, we find that the dominant states correspond to spin-singlet superconductivity,  spin-triplet pair-density-waves, and  spin-triplet superconductivity for fixed points $\textrm{FP}_{1,3}$, $\textrm{FP}_{2}$, and $\textrm{FP}_{4}$, respectively.
These provide valuable insights into the low-energy properties of rhombohedral trilayer graphene and analogous materials.
\end{abstract}


\maketitle


\section{Introduction}

The last two decades have witnessed remarkable progress in the study of Dirac materials
~\cite{Novoselov2005Nature,
Neto2009RMP,Kane2007PRL,Roy2009PRB,Moore2010Nature,Hasan2010RMP,Qi2011RMP,Sheng2012Book,
Bernevig2013Book,Huang2015PRX,Weng2015PRX,Hasan2015Science,Hasan2015NPhys,Ding2015NPhys}, including
the two-dimensional (2D) graphene~\cite{Neto2009RMP}, Weyl semimetals
~\cite{Neto2009RMP,Burkov2011PRL,Yang2011PRB,Savrasov2011PRB,Huang2015PRX,
Weng2015PRX,Hasan2015Science,Hasan2015NPhys,Ding2015NPhys}, moir'e graphene~\cite{Cao2018Nature,Cao2021Nature}
and other analogous compounds such as 2D semi-Dirac
materials~\cite{Hasegawa2006PRB,Pardo2009PRL,Katayama2006JPSJ,Dietl2008PRL,Delplace2010PRB,
Wu2014Expre,Uchoa1704.08780,Banerjee2009PRL,Saha2016PRB,Wang2018JPC,Wang2023arXiv,Fu2023arXiv,
Wang2022NPB,Zhai2021NPB,Dong2020PRB,Wang2019EPJB1,
Wang2019EPJB2,Wang2019JPC,Wang2018JPC,Wang2017PRB},
the quadratic-band-touching semimetals~\cite{Fradkin2009PRL,Vafek2014PRB1,Fu2023arXiv,Wang2022NPB,Zhai2021NPB,
Dong2020PRB,Wang2019EPJB1,Wang2019EPJB2,Wang2019JPC,Wang2018JPC,Wang2017PRB},
and Bernal bilayer graphene~\cite{Novoselov2006NPhys}.
Among these materials, rhombohedral trilayer graphene (RTG)~\cite{Zaliznyak2011NPhys,Koshino2009PRB,Zhang2010PRB},
a cousin of the graphene family, is of particular interest because it features the
bicubic band crossings at distinct corners of the hexagonal Brillouin zone~\cite{Zaliznyak2011NPhys}.
Such energy band yields distinctive low-energy excitations,
which contrast sharply with those of conventional
Dirac materials~\cite{Novoselov2005Nature,Neto2009RMP}. Besides, RTG is notable for being free from twist-related defects~\cite{Wilson2020PRR,Uri2020Nature,Mesple2021PRL,Artaud2018PRL,Cosma2014PRL,Bi2019PRB,Zhou2015PRB,Uchida2014PRB,Butz2014Nature}
and possessing a much simpler band structure compared with other materials~\cite{Koshino2009PRB,MacDonald2010PRB,Levitov2021CubicBand}.
As a result, RTG has garnered significant attention in recent years~\cite{Zhou2021Nature,Levitov2021CubicBand,Law2024arXiv,Levitov2024arXiv,
Roy2109.04466,MacDonald2203.12723,Weitz2024NPhys}.
Particularly, Zhou \emph{et al.}~\cite{Zhou2021Nature} discovered superconductivity in RTG. Subsequently,
Dong and Levitov proposed a simple model to describe this phenomenon by incorporating
the density-density interaction~\cite{Levitov2021CubicBand,Law2024arXiv,Levitov2024arXiv}.

It has been established that the fermion-fermion interactions play an important role in driving
critical behavior and shaping the low-energy properties of semimetals~\cite{Roy2004arXiv,Maiti2010PRB,Vafek2010PRB,Vafek2012PRB,Vafek2014PRB2,
Nandkishore2012NP,Nandkishore2013PRB,Chubukov2016PRX,Sur2016NJP,Roy2016Sci,Qi2016PRB,
Roy2016PRB,Nandkishore2017PRB,Roy2017PRB,Roy2017PRL,Benalcazar2017PRB,Wang2017PRB1,
Wang2018JPC,Wang2019JPC,Roy2018PRX1,Roy2018PRX2,Mandal2018PRB,Sur2019PRL,LBC-Wang,
Wang2020EPJB,Dong2020PRB,Zhai2021NPB,Szabo'2021PRB,Szabo'2021JHEP,Han2017PRB,Roy2016JHEP,
Cho2016Sci,Jian2017PRB,Tang2018Sci,Zhang2021PRB,Wang2023arXiv,Fu2023arXiv,Wang2023EPJB,Wang2022NPB,Wang2019EPJB1,
Wang2019EPJB2,Nandkishore2012NP}.
Considering the unique quasiparticle excitations in RTG, it is worth systematically examining the effects of fermion-fermion interactions on its low-energy physics. While important progress has been made in exploring RTG systems
~\cite{Zaliznyak2011NPhys, Koshino2009PRB, Zhang2010PRB, Levitov2021CubicBand, Zhou2021Nature, Law2024arXiv, Levitov2024arXiv, Roy2109.04466, MacDonald2203.12723, Weitz2024NPhys}, several intriguing issues remain to be addressed to improve our understanding. On one hand, to simplify the analysis,  fermion-fermion interactions are often partially taken into account, and several critical degrees of freedom may be neglected~\cite{Levitov2021CubicBand}. On the other hand, it has been observed that fermion-fermion interactions can drive distinct phase transitions in RTG, such as superconductivity and charge density waves~\cite{Levitov2021CubicBand}. Although mean-field theory may describe the basic results in the general case~\cite{Caines2022EDPS,Li2014JMAA,Mori2011PRE}, it fails to capture quantum fluctuations near potential instabilities induced by the fermion-fermion interactions in the low-energy regime. Such strong fluctuations often play an essential role in shaping low-energy properties~\cite{Roy2004arXiv,Maiti2010PRB,Vafek2010PRB,Vafek2012PRB,Vafek2014PRB2,
Nandkishore2012NP,Nandkishore2013PRB,Chubukov2016PRX,Sur2016NJP,Roy2016Sci,Qi2016PRB,
Roy2016PRB,Nandkishore2017PRB,Roy2017PRB,Roy2017PRL,Benalcazar2017PRB,Wang2017PRB1,
Wang2018JPC,Wang2019JPC,Roy2018PRX1,Roy2018PRX2,Mandal2018PRB,Sur2019PRL,Wang2020NPB,
Wang2020EPJB,Dong2020PRB,Zhai2021NPB,Szabo'2021PRB,Szabo'2021JHEP,Han2017PRB,Roy2016JHEP,
Cho2016Sci,Jian2017PRB,Tang2018Sci,Zhang2021PRB,Wang2023arXiv,Fu2023arXiv,Wang2023EPJB,Wang2022NPB,Wang2019EPJB1,Wang2019EPJB2,
LBC-Wang}. It is therefore of particular importance to examine the intricate fermion-fermion interactions in the low-energy regime and their impact on  low-energy properties of RTG.
For this purpose, in this paper, we not only consider the density-density interaction~\cite{Levitov2021CubicBand} but also
introduce all the other potential fermion-fermion interactions based on the RTG Hamiltonian. As a result,
sixteen fermion-fermion interactions are considered and then the Fierz
identity~\cite{Bian2023arVix,Boettcher2016PRB,Roy2017PRL,Bian2024arVix} is adopted
to derive six independent fermionic couplings, with which the effective action is established.
Starting from such an effective theory, we employ the powerful Wilsonian momentum-shell renormalization group (RG)~\cite{Wilson1975RMP,Polchinski9210046,Shankar1994RMP} to construct the energy-dependent coupled equations of all fermion-fermion interactions
after collecting all one-loop corrections. A detailed numerical analysis of these coupled evolutions reveals several
intriguing behavior in the low-energy regime.

At first, based on coupled RG equations, the interaction parameters exhibit strong energy-dependence and become increasingly
entangled with each other as lowering the energy scales, giving rise to distinct tendencies in the interaction space,
as summarized in Table~\ref{Table_I}. Specifically, the fermionic couplings are driven to a fixed point, namely $\textrm{FP}_{1}$~(\ref{Eq_4_2}), when
all interaction parameters are initially equal. In sharp contrast, when all interaction parameters
are independent and randomly assigned their initial values, they are attracted to three other fixed points,
denoted $\textrm{FP}_{2}$~(\ref{Eq_4_4}), $\textrm{FP}_{3}$~(\ref{Eq_4_8})
and $\textrm{FP}_{4}$~(\ref{Eq_4_9}). Subsequently, as the system approaches these fixed points,
we systematically investigate the potential instabilities and corresponding phase transitions
linked to specific symmetry breakings.
To this end, a number of fermion-bilinear source terms are introduced to describe the
candidate states associated with the phase transitions, as listed in Table~\ref{Table_II}. By evaluating and comparing the
susceptibilities of potential phases around these fixed points, we find that the most favorable states for the
$\textrm{FP}_{1,2,3,4}$ correspond to spin-singlet superconductivity ($\textrm{SC}_{1}$),
 spin-triplet pair-density-wave ($\textrm{PDW}_{2}$), $\textrm{SC}_{1}$,
and  spin-triplet superconductivity $\textrm{SC}_{3}$, respectively.
These findings provide insights for further studies of the RTG and analogous materials.

The rest of this paper is organized as follows. In Sec.~\ref{Sec_II}, we introduce the microscopic model and construct the effective theory. In Sec.~\ref{Sec_RG_eqations}, we derive the coupled RG equations of all fermion-fermion interaction parameters.
Next, in Sec.~\ref{Sec_fixed_points}, we systematically investigate the low-energy tendencies of interaction parameters and identity four distinct
fixed points in the low-energy regime. In Sec.~\ref{Sec_phase_transitions}, we examine the possible instabilities and associated
phase transitions near these fixed points. Finally, Sec.~\ref{Sec_summary} briefly
summarizes our basic conclusions.

\section{Microscopic model and effective action}\label{Sec_II}

On the basis of space-group symmetry and particle-hole symmetry constraints,
the microscopic noninteracting Hamiltonian of electrons in a rhombohedral trilayer graphene (RTG)   
can be expressed as follows~\cite{Levitov2021CubicBand,Koshino2009PRB,Zhang2010PRB}
\begin{eqnarray}
\textmd{H}_{0}=\sum_{\mathbf{p}}{\Psi_{\mathbf{p}} ^{\dag}}\mathcal{H}_{0}\Psi_{\mathbf{p}},\label{Eq_H0}
\end{eqnarray}
where the Hamiltonian density is written as~\cite{Levitov2021CubicBand,Koshino2009PRB,Zhang2010PRB}
\begin{eqnarray}
\mathcal{H}_{0}(\mathbf{p})=h_1 \tau_3 \otimes \sigma_1 +h_2 \tau_{0}\otimes\sigma_2+D\tau_{0}\otimes\sigma_3,\label{Eq_H0_2}
\end{eqnarray}
with $\mathbf{p}$ being the electronic momentum and two coefficients $h_1(\mathbf{p})=p_x^3 -3p_x p_y ^2$ plus $h_2(\mathbf{p})=\beta(3p_x^2 p_y-p_y^3)$
defined in the two-dimensional space along with the variable parameter $D$.
This indicates that the dynamical critical exponent of fermionic excitations equals
$z=3$, introducing distinct RG scalings for energy, momentum, and fermionic fields compared with those
of $z=1$ and $z=2$ and significantly impacting the RG equations as discussed in Sec.~\ref{Sec_RG_eqations}.
Here, the four-component spinor $\Psi=(\psi_{\emph{KA}},\psi_{\emph{KB}},\psi_{\emph{K}'\emph{A}},\psi_{\emph{K}'\emph{B}})^{T}$
characterizes the low-energy quasiparticle excitations. Both $\tau_i$ and $\sigma_i$ for $i=1,2,3$ ($i=0$ refers to the identity matrix)
are Pauli matrices corresponding to $\emph{K}$ and $\emph{K}'$ valley and sublattice (layer) bases, respectively~\cite{Levitov2021CubicBand}.

Subsequently, we introduce the short-range fermion-fermion interactions between low-energy excitations. According to the microscopic
model~(\ref{Eq_H0_2}), there exist sixteen distinct fermionic interactions that are explicitly listed as~\cite{Roy2004arXiv,Maiti2010PRB,Vafek2010PRB,Vafek2012PRB,Vafek2014PRB2,
Nandkishore2012NP,Nandkishore2013PRB,Chubukov2016PRX,Sur2016NJP,Roy2016Sci,Qi2016PRB,
Roy2016PRB,Nandkishore2017PRB,Roy2017PRB,Roy2017PRL,Benalcazar2017PRB,Wang2017PRB1,
Wang2018JPC,Wang2019JPC,Roy2018PRX1,Roy2018PRX2,Mandal2018PRB,Sur2019PRL,Wang2020NPB,
Wang2020EPJB,Dong2020PRB,Zhai2021NPB,Szabo'2021PRB,Szabo'2021JHEP,Han2017PRB,Roy2016JHEP,
Cho2016Sci,Jian2017PRB,Tang2018Sci,Zhang2021PRB,Wang2023arXiv,Fu2023arXiv,Wang2023EPJB,Wang2022NPB,Wang2019EPJB1,Wang2019EPJB2,LBC-Wang}
\begin{eqnarray}
&S_{\textrm{ff}}&=\sum_{i,j=0}^{3}\lambda_{ij}\prod_{\mu=1}^{3}\int\frac{d^{2}\mathbf{p}_{\mu}d\omega_{\mu}}{(2\pi)^{3}}
\Psi^{\dag}(i\omega_{1},\mathbf{p}_{1})\mathcal{C}_{ij}\Psi(i\omega_{2},\mathbf{p}_{2})\nonumber\\
&&\times\Psi^{\dag}(i\omega_{3},\mathbf{p}_{3})\mathcal{C}_{ij}\Psi(i\omega_{1}
\!+\!i\omega_{2}\!-\!i\omega_{3},\mathbf{p}_{1}+\mathbf{p}_{2}-\mathbf{p}_{3}),\label{Eq_ff_int}
\end{eqnarray}
where the vertex matrices $\mathcal{C}_{ij} \equiv \tau_{i}\otimes\sigma_{j}$ distinguish different fermion-fermion interactions, and
$\lambda_{ij}$ with $i,j=0,1,2,3$ quantify the corresponding strengths
of fermion-fermion couplings.
Fortunately, these fermion-fermion interactions are not independent from the point of view on linear algebra.
This motivates us to use the Fierz identity~\cite{Bian2023arVix,Boettcher2016PRB,Roy2017PRL,Bian2024arVix}
to reduce the number of interactions and simplify our calculations and analysis. In general,
the Fierz identity reads
\begin{eqnarray}
[\Psi^\dag(x)\mathcal{M}\Psi(x)]&&[\Psi^\dag(y)\mathcal{N}\Psi(y)]
=-\frac{1}{16}\sum_{ab}\mathrm{Tr}(\mathcal{M}\Gamma^{a}\mathcal{N}\Gamma^{b})\nonumber\\
&&\times[\Psi^\dag(x)\Gamma^{b}\Psi(y)]
[\Psi^\dag(y)\Gamma^{a}\Psi(x)],\label{Eq_Fierz}
\end{eqnarray}
where $\mathcal{M}$, $\mathcal{N}$, and $\Gamma$ correspond to the $4 \times 4$ matrices
and the matrix $\Gamma$ satisfies $(\Gamma_{a})^{\dag}=\Gamma^{a}=(\Gamma^{a})^{-1}$.
For convenience, we construct the interaction vector~\cite{Bian2023arVix,Boettcher2016PRB,Roy2017PRL,Szabo'2021PRB}
\begin{eqnarray}
&&\mathcal{B} = \{(\Psi^{\dag}\mathcal{C}_{00}\Psi)^{2},(\Psi^{\dag}\mathcal{C}_{01}\Psi)^{2},(\Psi^{\dag}\mathcal{C}_{02}\Psi)^{2},(\Psi^{\dag}\mathcal{C}_{03}\Psi)^{2},\nonumber\\
&&(\Psi^{\dag}\mathcal{C}_{10}\Psi)^{2},(\Psi^{\dag}\mathcal{C}_{11}\Psi)^{2},(\Psi^{\dag}\mathcal{C}_{12}\Psi)^{2},(\Psi^{\dag}\mathcal{C}_{13}\Psi)^{2},\nonumber\\
&&(\Psi^{\dag}\mathcal{C}_{20}\Psi)^{2},(\Psi^{\dag}\mathcal{C}_{21}\Psi)^{2},(\Psi^{\dag}\mathcal{C}_{22}\Psi)^{2},(\Psi^{\dag}\mathcal{C}_{23}\Psi)^{2},\nonumber\\
&&(\Psi^{\dag}\mathcal{C}_{30}\Psi)^{2},(\Psi^{\dag}\mathcal{C}_{31}\Psi)^{2},
(\Psi^{\dag}\mathcal{C}_{32}\Psi)^{2},(\Psi^{\dag}\mathcal{C}_{33}\Psi)^{2}\}.\label{Eq_vector}
\end{eqnarray}
With the help of Eq.~(\ref{Eq_Fierz}) and Eq.~(\ref{Eq_vector}), all sorts of the fermion-fermion interactions appearing in Eq.~(\ref{Eq_ff_int})
can be expressed by certain combinations of the other couplings, which yields
$\frac{1}{4}\sum_{j=1}^{16}{\mathfrak{F}_{ij}\mathcal{B}_{j}}=0$ with $\mathfrak{F}$ taking the form of
\begin{eqnarray}
\tiny{\mathfrak{F}=}
\left(\tiny{
  \begin{smallmatrix}
    5 & 1 & 1 & 1 & 1 & 1 & 1 & 1 & 1 & 1 & 1 & 1 & 1 & 1 & 1 & 1 \\
    1 & 5 & -1 & -1 & 1 & 1 & -1 & -1 & 1 & 1 & -1 & -1 & 1 & 1 & -1 & -1 \\
    1 & -1 & 5 & -1 & 1 & -1 & 1 & -1 & 1 & -1 & 1 & -1 & 1 & -1 & 1 & -1 \\
    1 & -1 & -1 & 5 & 1 & -1 & -1 & 1 & 1 & -1 & -1 & 1 & 1 & -1 & -1 & 1 \\
    1 & 1 & 1 & 1 & 5 & 1 & 1 & 1 & -1 & -1 & -1 & -1 & -1 & -1 & -1 & -1 \\
    1 & 1 & -1 & -1 & 1 & 5 & -1 & -1 & -1 & -1 & 1 & 1 & -1 & -1 & 1 & 1 \\
    1 & -1 & 1 & -1 & 1 & -1 & 5 & -1 & -1 & 1 & -1 & 1 & -1 & 1 & -1 & 1 \\
    1 & -1 & -1 & 1 & 1 & -1 & -1 & 5 & -1 & 1 & 1 & -1 & -1 & 1 & 1 & -1 \\
    1 & 1 & 1 & 1 & -1 & -1 & -1 & -1 & 5 & 1 & 1 & 1 & -1 & -1 & -1 & -1 \\
    1 & 1 & -1 & -1 & -1 & -1 & 1 & 1 & 1 & 5 & -1 & -1 & -1 & -1 & 1 & 1 \\
    1 & -1 & 1 & -1 & -1 & 1 & -1 & 1 & 1 & -1 & 5 & -1 & -1 & 1 & -1 & 1 \\
    1 & -1 & -1 & 1 & -1 & 1 & 1 & -1 & 1 & -1 & -1 & 5 & -1 & 1 & 1 & -1 \\
    1 & 1 & 1 & 1 & -1 & -1 & -1 & -1 & -1 & -1 & -1 & -1 & 5 & 1 & 1 & 1 \\
    1 & 1 & -1 & -1 & -1 & -1 & 1 & 1 & -1 & -1 & 1 & 1 & 1 & 5 & -1 & -1 \\
    1 & -1 & 1 & -1 & -1 & 1 & -1 & 1 & -1 & 1 & -1 & 1 & 1 & -1 & 5 & -1 \\
    1 & -1 & -1 & 1 & -1 & 1 & 1 & -1 & -1 & 1 & 1 & -1 & 1 & -1 & -1 & 5 \\
  \end{smallmatrix}}
\right).
\end{eqnarray}
After simplifying this matrix, we find that $\mathrm{rankNull}(\mathfrak{F})=6$, and implying that
only six of the fermion-fermion interactions in Eq.~(\ref{Eq_ff_int}) are independent~\cite{Bian2023arVix,Boettcher2016PRB,Roy2017PRL,Szabo'2021PRB}.
Consequently, it is convenient to select six representative types of interactions to establish the interacting terms. Following the strategy in
~\cite{Bian2023arVix,Boettcher2016PRB,Roy2017PRL,Szabo'2021PRB}, we reformulate the effective fermion-fermion interactions as follows
\begin{eqnarray}
&&S_{\textrm{int}}=\sum_{i=1}^{6}\lambda_{i}\prod_{\mu=1}^{3}\int\frac{d^{2}\mathbf{p}_{\mu}d\omega_{\mu}}{(2\pi)^{3}}
\Psi^{\dag}(i\omega_{1},\mathbf{p}_{1})\mathcal{V}_{i}\Psi(i\omega_{2},\mathbf{p}_{2})\nonumber\\
&&\times\Psi^{\dag}(i\omega_{3},\mathbf{p}_{3})\mathcal{V}_{i}\Psi(i\omega_{1}
+i\omega_{2}-i\omega_{3},\mathbf{p}_{1}+\mathbf{p}_{2}-\mathbf{p}_{3}),\label{Eq_S_int}
\end{eqnarray}
where $\mathcal{V}_{i}$ stand for $\mathcal{C}_{02}$, $\mathcal{C}_{03}$, $\mathcal{C}_{10}$,
$\mathcal{C}_{12}$, $\mathcal{C}_{20}$, and $\mathcal{C}_{22}$, while $\lambda_{i}$ represent
the corresponding strengths of fermion-fermion interactions with $i$ running from 1 to 6.

To proceed, considering the free Hamiltonian in Eqs.~(\ref{Eq_H0})-(\ref{Eq_H0_2}) together with the interacting terms~(\ref{Eq_S_int}),
we obtain effective theory
\begin{eqnarray}
S_{\textrm{eff}}
&=&\!\!\!\int\frac{d^{2}\mathbf{p}d\omega}{(2\pi)^{3}}\Psi^{\dag}(i\omega ,\mathbf{p})
(-i\omega+h_1 \tau_3 \otimes \sigma_1 +h_2 \tau_{0}\otimes\sigma_2\nonumber\\
&&+D\tau_{0}\otimes\sigma_3)\Psi (i\omega ,\mathbf{p})
+S_{\textrm{int}},\label{Eq_S_eff}
\end{eqnarray}
where $\omega$ denotes the frequency. From the effective theory~(\ref{Eq_S_eff}), the free fermionic propagator can be
obtained as
\begin{eqnarray}
G_{0}(i\omega ,\mathbf{p})=\frac{1}{-i\omega +h_1 \mathcal{C}_{31} +h_2 \mathcal{C}_{02} +D \mathcal{C}_{03}},\label{Eq_G_0}
\end{eqnarray}
which is crucial for evaluating one-loop corrections in Sec.~\ref{Sec_RG_eqations}.
In the forthcoming sections,  we focus on the effective action~(\ref{Eq_S_eff}) and investigate
the critical physical behavior induced by fermion-fermion interactions in RTG.

\section{RG flow equations}\label{Sec_RG_eqations}

To investigate the effects of fermion-fermion interactions on low-energy physics, we adopt the Wilsonian momentum-shell RG~\cite{Wilson1975RMP,Polchinski9210046,Shankar1994RMP}, with which the energy-dependent interaction parameters in the
effective action~(\ref{Eq_S_eff}) that encapsulate low-energy physics can be derived. According to the RG formalism, we treat the free term of the effective action~(\ref{Eq_S_eff}) as an initial fixed point that remains invariant under RG transformations~\cite{Wilson1975RMP,Polchinski9210046,Shankar1994RMP}. Using this restriction, the RG rescaling transformations for momenta, energy, and fermionic fields are given by~\cite{Shankar1994RMP,Huh2008PRB,She2010PRB,
She2015PRB,Wang2011PRB,Wang2022NPB,Kim2008PRB,Wang2013PRB,Wang2017PRB2,Wang2014PRD,Wang2015PRB,Wang2022SST}
\begin{eqnarray}
p_{x}&\longrightarrow& p_{x}'e^{-l},\label{Eq_scaling-1}\\
p_{y}&\longrightarrow& p_{y}'e^{-l},\label{Eq_scaling-2}\\
\omega&\longrightarrow& \omega'e^{-3l},\label{Eq_scaling-3}\\
\Psi(i\omega,\mathbf{p})&\longrightarrow&\Psi^{'}(i\omega',\mathbf{p}')e^{\frac{1}{2}\int dl(8-\eta_{f})},\label{Eq_scaling-4}
\end{eqnarray}
where the parameter $\eta_{f}$ denotes the potential anomalous dimension of fermionic spinor.
Next, we go beyond the tree level and evaluate all one-loop corrections by integrating out
the high-momentum modes within the shell $b\Lambda < k < \Lambda$, where $\Lambda$ represents the cutoff energy scale
and the parameter $b$ is defined as $b = e^{-l}$, with the running scale $l > 0$ indicating the variation in energy scale.
After lengthy but straightforward calculations,
the one-loop corrections are derived and provided by Eqs.~(\ref{Eq_Appendix_1})-(\ref{Eq_Appendix_2}) in Appendix~\ref{Appendix_1L-corrections}, which
indicates $\eta_f=0$ to the one-loop level.
Subsequently, employing these one-loop corrections in tandem with the RG transformation scalings~(\ref{Eq_scaling-1})-(\ref{Eq_scaling-4})
yields the one-loop coupled RG flow equations for all parameters in the effective action~(\ref{Eq_S_eff})~\cite{Shankar1994RMP,Kim2008PRB,Huh2008PRB,She2010PRB,She2015PRB,Wang2011PRB},
\begin{widetext}
\begin{eqnarray}
\frac{dD}{dl}
&=&3D,\label{Eq_3_6}\\
\frac{d\lambda_{1}}{dl}
&=&\lambda_{1}+2\lambda_{1}[(\lambda_{1}-\lambda_{2}+\lambda_{3}+\lambda_{4}+\lambda_{5}
+\lambda_{6})(\mathbb{A+B-C+D})-4\lambda_{1}(\mathbb{A+B-C+D})],\\\label{Eq_3_7}
\frac{d\lambda_{2}}{dl}
&=&\lambda_{2}-2\lambda_{2}[(\lambda_{1}-\lambda_{2}-\lambda_{3}+\lambda_{4}-\lambda_{5}
+\lambda_{6})(\mathbb{A+B+C-D})-4\lambda_{2}(\mathbb{A+B+C-D})]\nonumber\\
&&-2(2\lambda_{3}\lambda_{6}+2\lambda_{4}\lambda_{5})\mathbb{B}-2(\lambda_{1}^{2}+\lambda_{2}^{2}
+\lambda_{3}^{2}+\lambda_{4}^{2}+\lambda_{5}^{2}+\lambda_{6}^{2})\mathbb{D}-4\lambda_{1}\lambda_{2}\mathbb{C},\\\label{Eq_3_8}
\frac{d\lambda_{3}}{dl}
&=&\lambda_{3}+2\lambda_{3}[(\lambda_{1}+\lambda_{2}+\lambda_{3}+\lambda_{4}-\lambda_{5}
-\lambda_{6})(\mathbb{A+B-C-D})-4\lambda_{3}(\mathbb{A+B-C-D})]\nonumber\\
&&-2(2\lambda_{2}\lambda_{6}\mathbb{B}+2\lambda_{2}\lambda_{3}\mathbb{D}),\\\label{Eq_3_9}
\frac{d\lambda_{4}}{dl}
&=&\lambda_{4}+2\lambda_{4}[(\lambda_{1}-\lambda_{2}+\lambda_{3}+\lambda_{4}-\lambda_{5}
-\lambda_{6})(\mathbb{A-B-C+D})-4\lambda_{4}(\mathbb{A-B-C+D})],\\\label{Eq_3_10}
\frac{d\lambda_{5}}{dl}
&=&\lambda_{5}+2\lambda_{5}[(\lambda_{1}+\lambda_{2}-\lambda_{3}-\lambda_{4}+\lambda_{5}
+\lambda_{6})(\mathbb{A+B-C-D})-4\lambda_{5}(\mathbb{A+B-C-D})]\nonumber\\
&&-2(2\lambda_{2}\lambda_{4}\mathbb{B}+2\lambda_{2}\lambda_{5}\mathbb{D}),\\\label{Eq_3_11}
\frac{d\lambda_{6}}{dl}
&=&\lambda_{6}+2\lambda_{6}[(\lambda_{1}-\lambda_{2}-\lambda_{3}-\lambda_{4}+\lambda_{5}
+\lambda_{6})(\mathbb{A-B-C+D})-4\lambda_{6}(\mathbb{A-B-C+D})],\label{Eq_3_12}
\end{eqnarray}
\end{widetext}
where the coefficients $\mathbb{A}-\mathbb{D}$ are explicitly designated in Appendix~\ref{Appendix_1L-corrections}.
These coupled RG equations~(\ref{Eq_3_6})-(\ref{Eq_3_12}) are intimately entangled and govern the
physics in the low-energy regime. Under the constraints of their coupled evolutions, these interaction parameters flow toward
certain fixed points at the lowest-energy limit. These so-called fixed points exist in the parameter spaces spanned by
the fermion-fermion interaction
parameters~\cite{Maiti2010PRB,Vojta2003RPP,Roy2018PRX2,Vafek2012PRB,Vafek2014PRB2,
Chubukov2012ARCMP,Chubukov2016PRX,Nandkishore2012NP,Fu2023arXiv}.
Next, we investigate the potential fixed points arising from the RG equations in Sec.~\ref{Sec_fixed_points} and then
move to Sec.~\ref{Sec_phase_transitions} to examine the underlying instabilities and leading phase transitions near such fixed points.

\begin{figure}[htbp]
\centering
\includegraphics[width=3.0in]{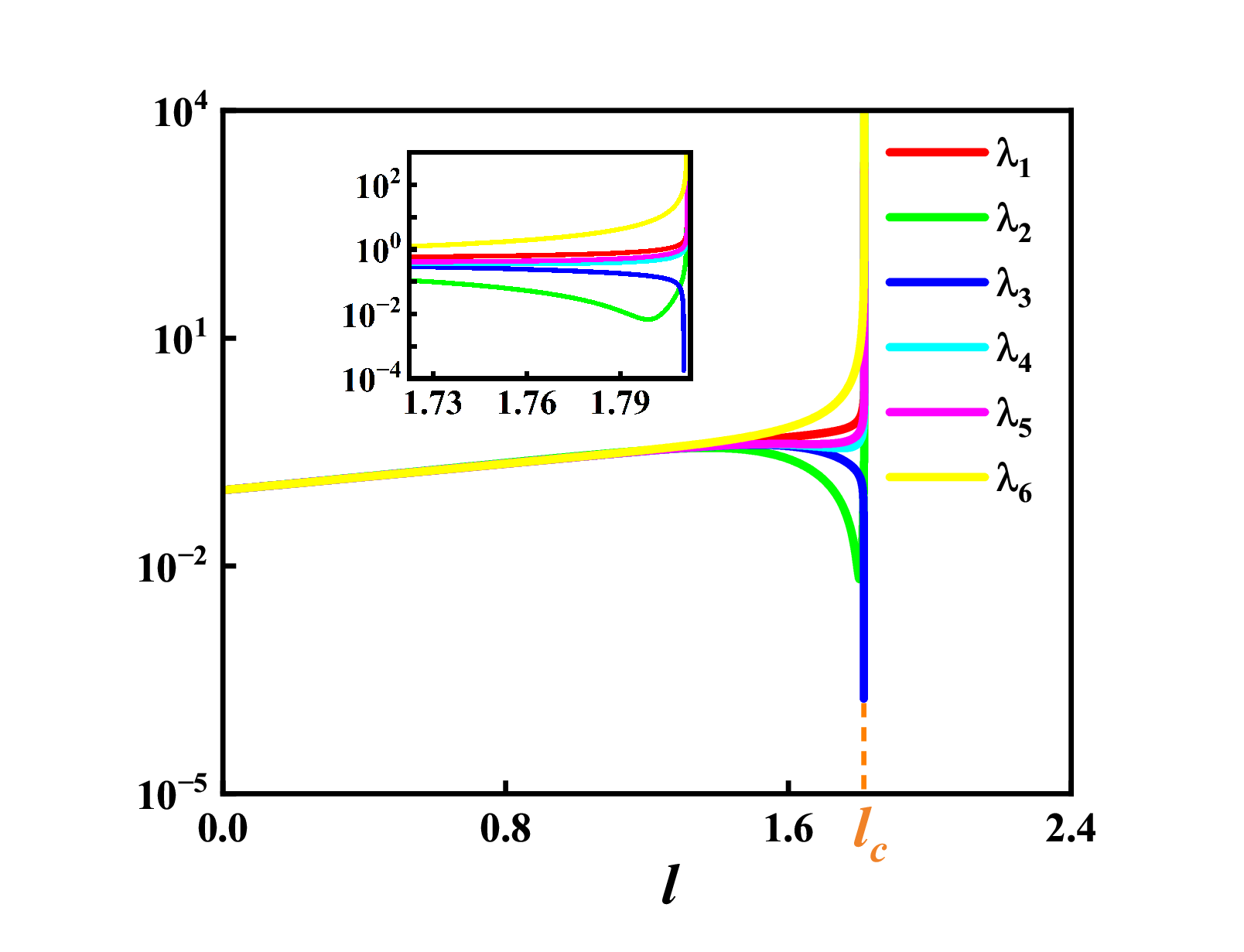}
\caption{(Color online) Energy-dependent flows of the fermion-fermion interaction parameters with a representative
initial condition, namely $D_{0}=10^{-1}$ and $\lambda_{i0}=10^{-1}(i=1-6)$ (the basic results are insusceptible
to the concrete initial values). Inset: the enlarged regime around the critical energy scale.}\label{Fig_4}
\end{figure}

\section{Fixed points of interaction parameters}\label{Sec_fixed_points}

Starting from the coupled RG equations~(\ref{Eq_3_6})-(\ref{Eq_3_12}), we now examine the low-energy tendencies of
the interaction parameters. Performing the numerical analysis of these RG equations
reveals that all interaction parameters are strongly energy-dependent and intimately entangled as the energy scale decreases.
As a consequence, they exhibit various interesting evolutions,
and, in particular, their low-energy behavior is susceptible to the initial conditions, which we
address in the following.

\subsection{Special case}\label{Sec_IV_A}

At the outset, we consider a special case, in which all the interaction parameters are
initially equal. After numerical analysis, we
present the basic tendencies of interaction parameters for this situation in Fig.~\ref{Fig_4} with a representative initial condition.
From Fig.~\ref{Fig_4}, we notice that several interaction parameters increase with decreasing the energy scale and
flow towards divergence at a critical energy scale $l=l_c$, beyond which the RG equations become invalid.

In order to make our analysis under control, we adopt the strategy in Refs.~\cite{Vafek2012PRB,Vafek2014PRB2,Roy2018PRX2,Fu2023arXiv} by introducing rescaled interaction parameters, $\lambda_i\rightarrow\lambda_i/\lambda_m$ where $\lambda_m=\max\{|\lambda_{i}|\}$ with $i=1-6$.
On this basis, we identify the fixed point in the interaction-parameter space to characterize the final behavior of these
couplings. Specifically, such a fixed point is represented as coordinates in the interaction-parameter space at the critical
energy scale, namely
\begin{eqnarray}
\textrm{FP}|_{l=l_c}\equiv\left.\left(\frac{\lambda_{1}}{\lambda_m},\frac{\lambda_{2}}{\lambda_m},\frac{\lambda_{3}}{\lambda_m},
\frac{\lambda_{4}}{\lambda_m},\frac{\lambda_{5}}{\lambda_m},\frac{\lambda_{6}}{\lambda_m}\right)\right|_{l=l_c}.\label{Eq_4_1}
\end{eqnarray}
Figure~\ref{Fig_4} shows that $\lambda_6$ dominates the other parameters in this special case. It is therefore
appropriate to rescale all interaction parameters using this very parameter.

\begin{figure}[htbp]
\centering
\includegraphics[width=3.0in]{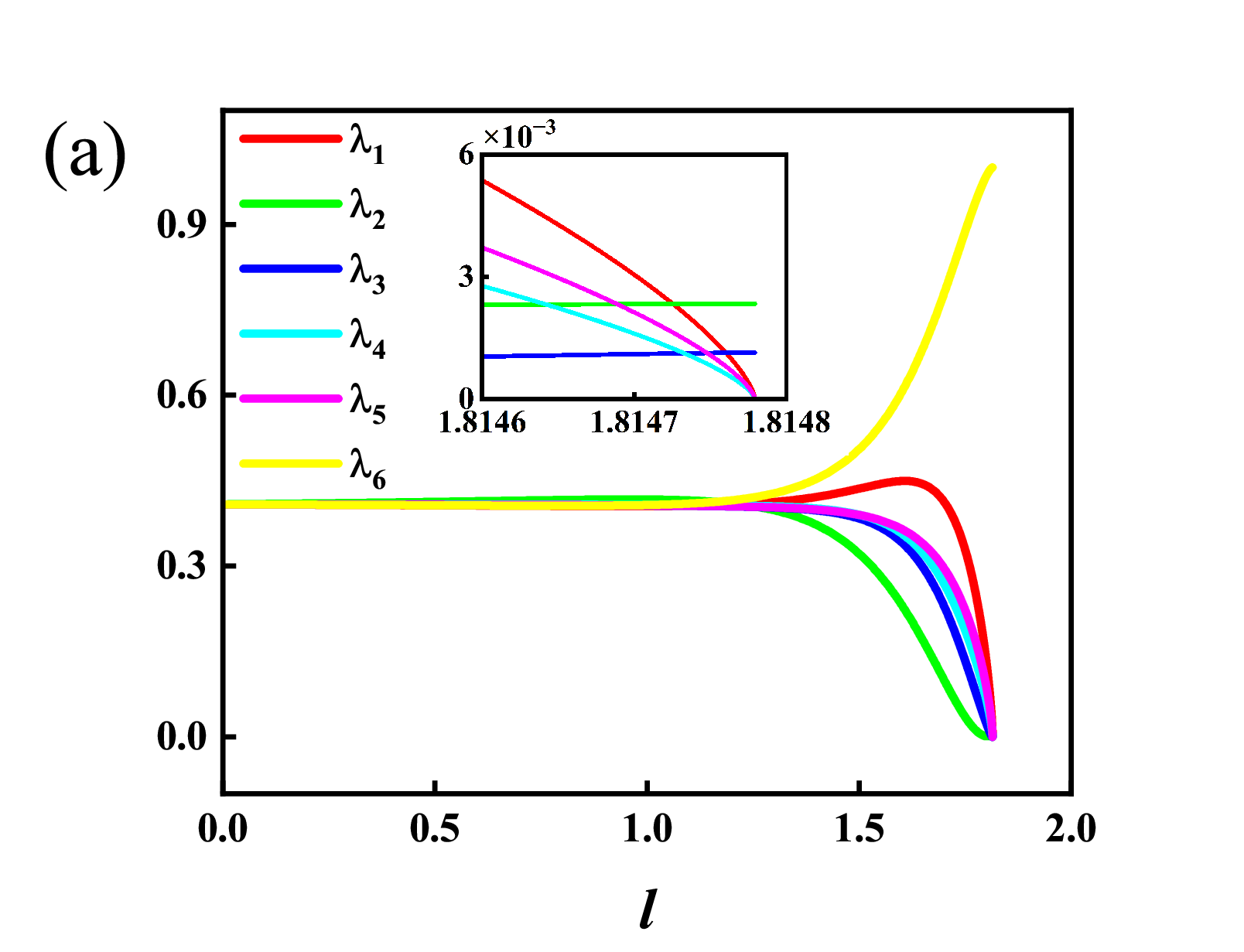}
\includegraphics[width=3.0in]{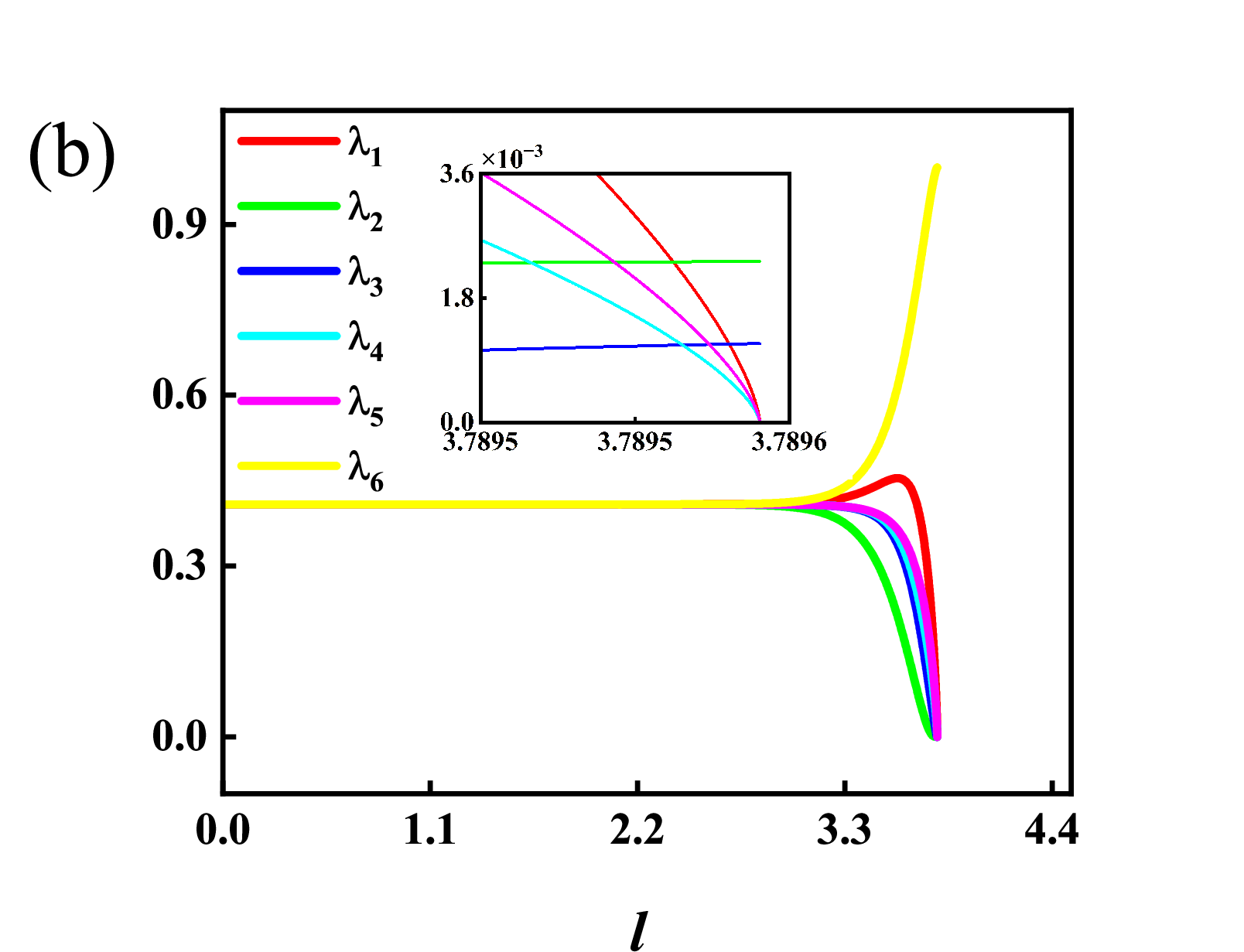}
\caption{(Color online) Energy-dependent flows of the rescaled fermion-fermion interaction parameters for $D_{0}=10^{-1}$
plus (a) $\lambda_{i0}=10^{-1}$,  and (b) $\lambda_{i0}=10^{-7}$ with $(i=1-6)$ (the basic results are insusceptible
to the concrete values of $\lambda_{i0}$ from $\lambda_{i0}=10^{-7}$ to $10^{-1}$). (Inset) The enlarged regime around
the critical energy scale.}\label{Fig_5}
\end{figure}

Under this condition, while fixing an initial value of $D$ ($D_0$), we vary the initial values of interaction parameters
from $\lambda_{i0}=10^{-7}$ to $10^{-1}$  (for convenience,
from now on, $\lambda_{i0}$ and $D_0$ denote the initial
values of $\lambda_i$ and $D$, respectively) and find the qualitative results remain similar. Besides, the basic results of
interaction couplings are insensitive to the variations in $D_0$. Thus, without loss of generality, we select two representative $\lambda_{i0}$ values and a
$D_0$ to construct Fig.~\ref{Fig_5} for this special case. From Fig.~\ref{Fig_5}(a), it is evident that $\lambda_6$
emerges as the dominant parameter at sufficiently low energy scales,
which is insusceptible to the concrete value of $\lambda_{i0}$ with comparing Fig.~\ref{Fig_5}(a) and Fig.~\ref{Fig_5}(b).
In addition, because of the strong fluctuations, the differences among these interactions are manifestly enlarged as the system approaches  the critical energy scale. Based on the energy-dependent evolutions in Fig.~\ref{Fig_5} and the definition of fixed point~(\ref{Eq_4_1}),
we obtain the fixed point at $l=l_c$ for the special case, denominated as $\textrm{FP}_{1}$,
\begin{eqnarray}
\textrm{FP}_{1}|_{l=l_c}\approx(0,0.002,0,0,0,1)\label{Eq_4_2}.
\end{eqnarray}
As noted earlier, the qualitative conclusions are robust enough against the variations of both $\lambda_{i0}$ and $D_0$,
as long as the restriction of special case is satisfied, namely $\lambda_{i0}$ being equal.
Consequently, the $\textrm{FP}_{1}$ is the only fixed point and in principle expected to govern the low-energy
physics in this special case.

\begin{figure*}[htbp]
\centering
\includegraphics[width=2.5in]{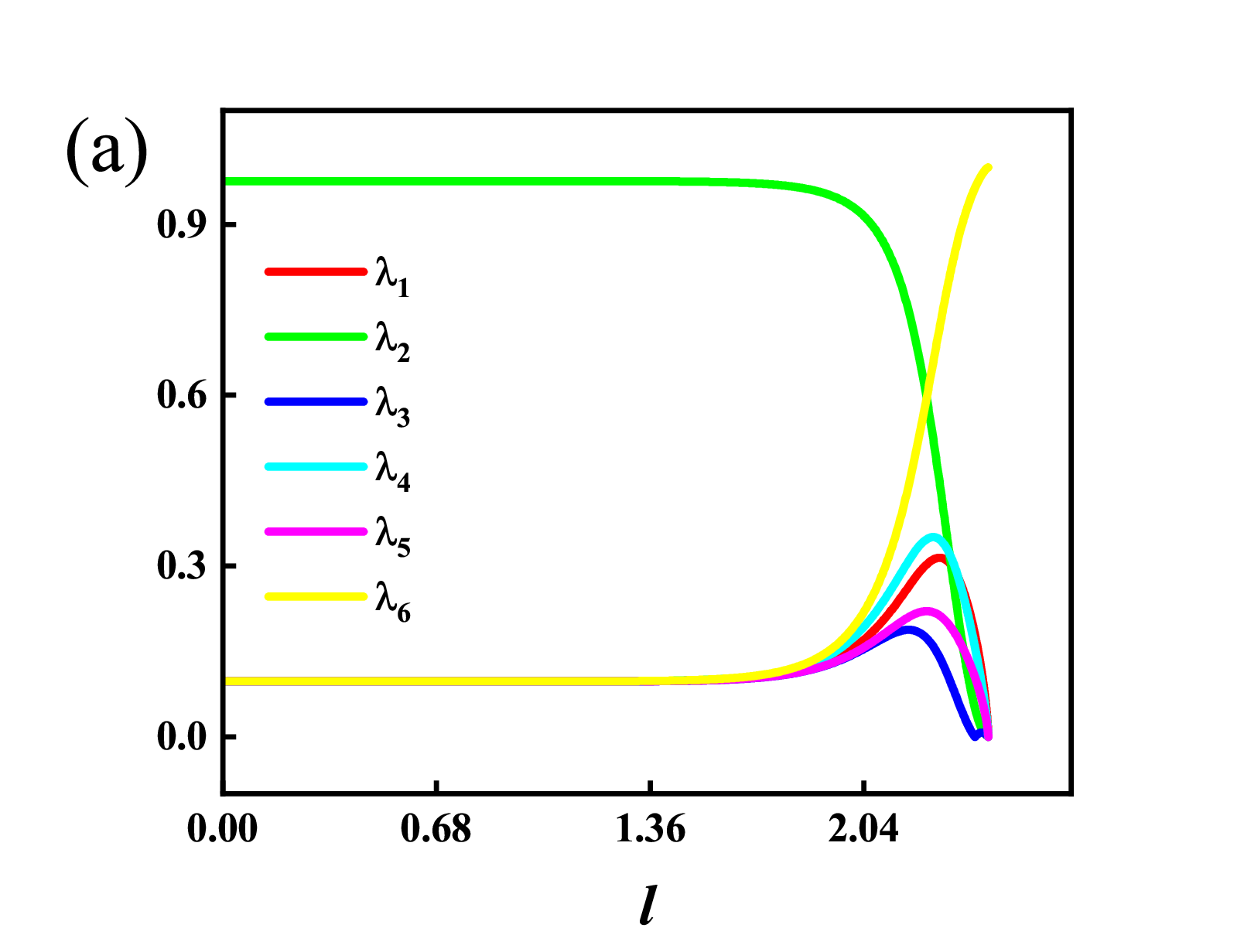}
\includegraphics[width=2.5in]{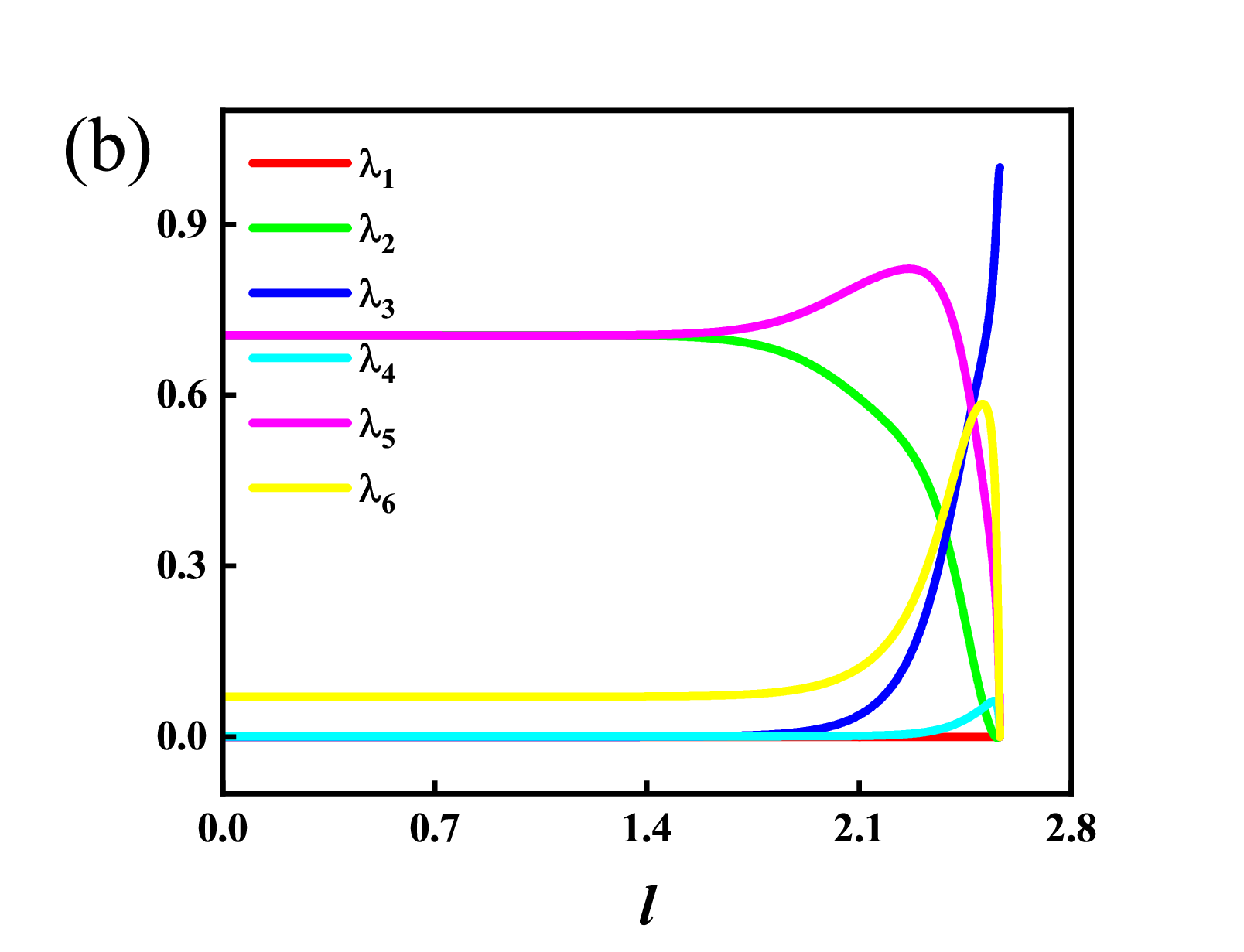}\\ \vspace{-0.5cm}
\includegraphics[width=2.5in]{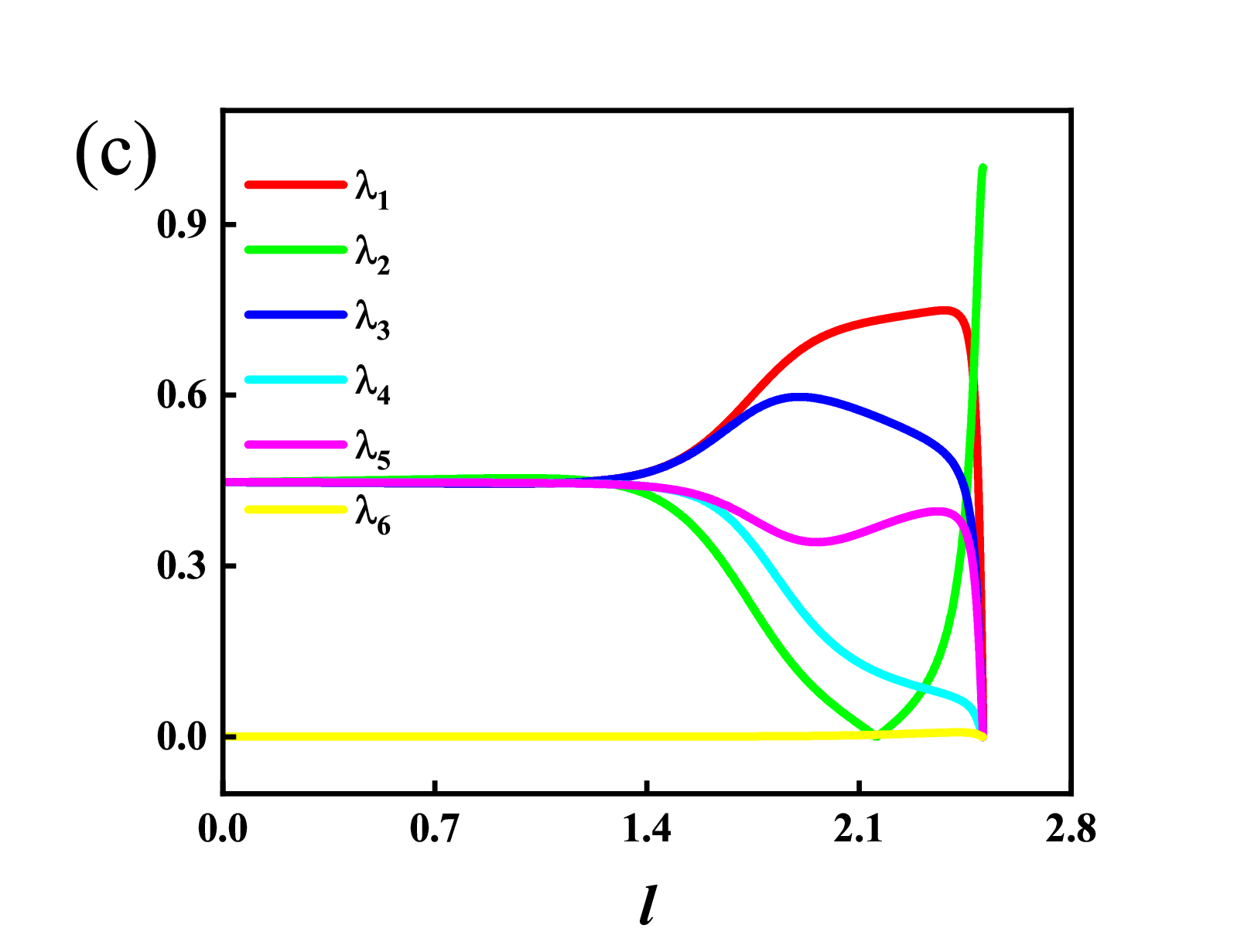}
\includegraphics[width=2.5in]{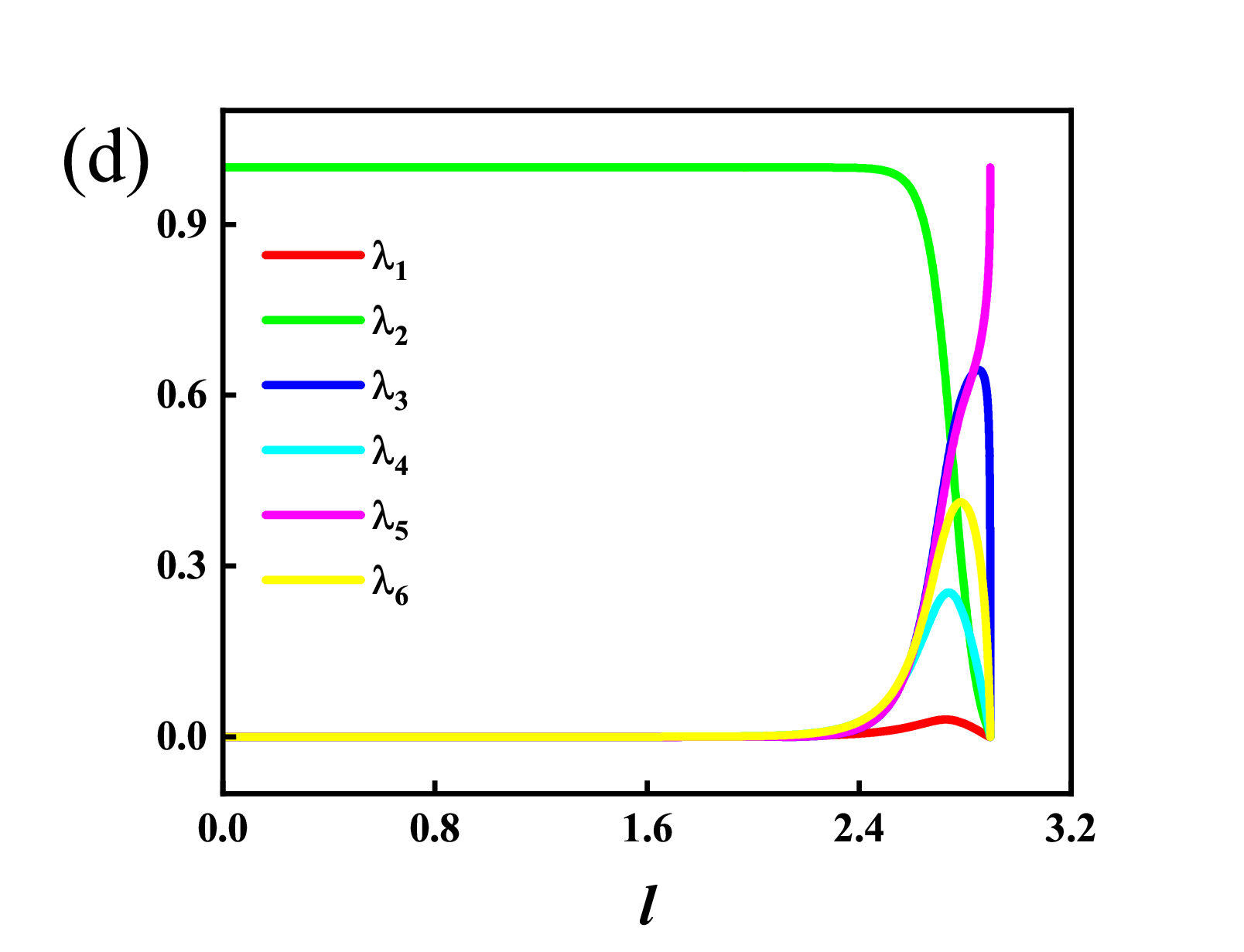}
\caption{(Color online) Energy-dependent evolutions of the rescaled interaction parameters with $D_{0}=10^{-1}$ plus
(a) $\lambda_{i0}=10^{-3}(i=1,3,4,5,6), \lambda_{20}=10^{-2}$ flowing towards $\textrm{FP}_{1}$,
(b) $D_{0}=10^{-1}$, $\lambda_{10}=10^{-7}, \lambda_{20}=10^{-2}, \lambda_{30}=10^{-7}, \lambda_{40}=10^{-5},
\lambda_{50}=10^{-2}, \lambda_{60}=10^{-3}$ flowing towards $\textrm{FP}_{2}$, (c) $D_{0}=10^{-1}$,
$\lambda_{i0}=10^{-1}(i=1,2,3,4,5), \lambda_{60}=10^{-4}$ flowing towards  $\textrm{FP}_{3}$, and
(d) $D_{0}=10^{-1}$, $\lambda_{i0}=10^{-5}(i=1,3,4,5,6), \lambda_{20}=10^{-1}$ flowing towards $\textrm{FP}_{4}$.}\label{Fig_6}
\end{figure*}

\subsection{General case}\label{Sec_IV_B}

Subsequently, let us consider the general case, in which all six types of interaction parameters randomly
take their own starting values.

After performing the similar numerical analysis in Sec.~\ref{Sec_IV_A} with varying $\lambda_{i0}$ from $10^{-7}$ to $10^{-1}$ for the
general case,  Fig.~\ref{Fig_6} shows that the tendencies of fermion-fermion interactions,
in sharp contrast to those of the special case, are heavily dependent on their initial values.
It is noteworthy that, compared to the sole leading interaction $\lambda_6$ in the special case,
four candidate couplings, namely $\lambda_{2}$, $\lambda_{3}$, $\lambda_{5}$, and $\lambda_{6}$ can become the dominant
interaction, depending on the initial conditions.

While $\lambda_6$ is dominant as depicted in Fig.~\ref{Fig_6}(a), the system is again attracted to the $\textrm{FP}_{1}$.
However, three new fixed points can be established once each of the other three couplings takes the leading place.
As shown in Fig.~\ref{Fig_6}(b), the leading $\lambda_3$ yields the second fixed point, which is defined as
$\textrm{FP}_{2}$ and takes the form
\begin{eqnarray}
\textrm{FP}_{2}|_{l=l_c}\approx(0,0.002,-1,0,0,0).\label{Eq_4_4}
\end{eqnarray}
It is worth highlighting that this fixed point is obtained by satisfying a relatively strict initial condition, which is named the
$\textrm{FP}_{2}\mathrm{-Condition}$ for further reference. After careful numerical analysis, $\textrm{FP}_{2}\mathrm{-Condition}$ is constructed as
\begin{eqnarray}
&&\textrm{FP}_{2}\mathrm{-Condition}: \lambda_{10}>0,\,\,\lambda_{20}>10^{-2},\,\,\lambda_{30}\in(0,10^{-5})\nonumber\\
&&\,\,\,\,\,\,\,\lambda_{40}\in(0,10^{-5}),\,\lambda_{50}>10^{-2},\,\lambda_{60}\in(10^{-5},10^{-3}).\label{Eq_4_3}
\end{eqnarray}
Then, paralleling the similar studies gives rises to another two fixed points as shown in Fig.~\ref{Fig_6}(c) and Fig.~\ref{Fig_6}(d),
\begin{eqnarray}
\textrm{FP}_{3}|_{l=l_c}&\approx&(0,-1,0,0,0,0),\label{Eq_4_8}\\
\textrm{FP}_{4}|_{l=l_c}&\approx&(0,0.002,0,0,-1,0),\label{Eq_4_9}
\end{eqnarray}
which correspond to the $\lambda_2$-dominant and $\lambda_5$-dominant situations, respectively.

In order to seek the initial conditions for emergence of such two fixed points, it is useful to introduce three auxiliary variables,
\begin{eqnarray}
\kappa &\equiv& \frac{\lambda_{i0}}{\lambda_{60}}\, (i = 1,2,3,4,5),\\\label{Eq_4_5}
\eta &\equiv& \frac{\lambda_{j0}}{\lambda_{20}}\, (j = 1,3,4,5,6),\\\label{Eq_4_6}
\zeta &\equiv& \frac{\lambda_{k0}}{\lambda_{50}}\, (k = 1,2,3,4,6).\label{Eq_4_7}
\end{eqnarray}
With the help of these three parameters, we find that the initial conditions that realize $\textrm{FP}_{3}$ and
$\textrm{FP}_{4}$ can be compactly expressed. Concretely, our numerical analysis indicates that
$\lg\ \kappa >2$\,\,\,\,(or\,\,\,\,$\lg\ \zeta >-2)$ generates $\textrm{FP}_{3}$, whereas
$\textrm{FP}_{4}$ can be induced at $\lg\ \eta < -4$.
Additionally, it is of particular interest to emphasize that although the $\textrm{FP}_{1}$ can be reached for the general case as
displayed in Fig.~\ref{Fig_6}(a),
its initial condition is more stringent than its special case counterpart. Borrowing the auxiliary variables leads to
the explicit initial relationship for $\textrm{FP}_{1}$ as
$\lg\ \kappa < 2$\,\,\,\,or\,\,\,\,$\lg\ \eta > -4$\,\,\,\,or\,\,\,\,$\lg\ \zeta < -2$.
Based on these discussions, we present Table~\ref{Table_I} to recapitulate the initial conditions that are required to
induce all four distinct types of fixed points for the general case.

\begin{table}
\caption{Collections of required initial conditions for the general case to realize four distinct types of fixed points.}\vspace{0.3cm}
\centering{
\renewcommand\arraystretch{2}
\begin{tabular}{cc}
\hline
\hline
Fixed points & Conditions \\
\hline
$\textrm{FP}_{1}$ & $\lg\ \kappa < 2$\,\,\,\,or\,\,\,\,$\lg\ \eta > -4$\,\,\,\,or\,\,\,\,$\lg\ \zeta < -2$\\
$\textrm{FP}_{2}$ & $\textrm{FP}_{2}\mathrm{-Condition}$~(\ref{Eq_4_3}) \\
$\textrm{FP}_{3}$ & $\lg\ \kappa >2$\,\,\,\,or\,\,\,\,$\lg\ \zeta >-2$\\
$\textrm{FP}_{4}$ & $\lg\ \eta < -4$\\
\hline
\hline
\end{tabular}
}\label{Table_I}
\end{table}

To wrap up, a comprehensive analysis of the coupled RG equations~(\ref{Eq_3_6})-(\ref{Eq_3_12})
indicates that four distinct types of fixed points exist as summarized in Table~\ref{Table_I},
owing to the intimate competition among all types of interaction parameters in the low-energy regime.
Generally, the critical behavior can be induced as these fixed
points are approached~\cite{Vafek2012PRB,Vafek2014PRB2,Roy2018PRX2,Fu2023arXiv}, which
will be addressed in the forthcoming Sec.~\ref{Sec_phase_transitions}.

\section{Potential phase transitions}\label{Sec_phase_transitions}

In principle, the fixed points with divergent couplings are indicative of instabilities exactly at the
critical energy scales because of ferocious quantum fluctuations~\cite{Maiti2010PRB,Vafek2012PRB,Vojta2003RPP,
Chubukov2012ARCMP,Vafek2014PRB2,Halboth2000RPL,Halboth2000RPB,Chubukov2016PRX,Nandkishore2012NP,Wang2020NPB,Roy2018PRX2,Eberlein2014PRB}.
This, in turn, induces certain phase transitions
with symmetry breakings~\cite{Vafek2014PRB2, Maiti2010PRB,Halboth2000RPL,
Nandkishore2012NP,Chubukov2016PRX,Roy2018PRX2}.
It is therefore particularly important to examine whether instabilities and phase
transitions occur around the fixed points obtained in the previous section, and to determine the dominant
states after experiencing
the potential phases.

\begin{table*}
\caption{Eight different types of potential states driven by fermion-fermion interactions in the
RTG systems~\cite{Levitov2021CubicBand}, which are correspondingly associated with fermion-bilinear source terms defined
in Eq.~(\ref{Eq_5_1}).}\vspace{0.3cm}
\centering{
\renewcommand\arraystretch{2}
\begin{tabular}{ccccc}
\hline
\hline
 Candidate states & \hspace{0.05cm} Vertex matrices \hspace{1.25cm}& \hspace{0.35cm}spin \hspace{1.25cm}& \hspace{1.25cm}spatial symmetry \hspace{1.25cm}&\hspace{0.35cm} irrep \\
\hline
 $\textrm{SC}_{1}$ & \hspace{0.35cm} $\mathcal{M}_{1}=\sigma_{1}\otimes\tau_{1}$ \hspace{0.35cm} & \hspace{0.25cm} spin singlet & \hspace{0.8cm} no symmetries broken &\hspace{0.55cm} $A_{1,\Gamma}^{+}$\\
 $\textrm{SC}_{2}$ & \hspace{0.35cm} $\mathcal{M}_{2}=\sigma_{1}\otimes\tau_{2}$ \hspace{0.35cm} & \hspace{0.25cm} spin triplet &\hspace{0.8cm} no symmetries broken &\hspace{0.55cm} $A_{1,\Gamma}^{-}$\\
 $\textrm{SC}_{3}$ & \hspace{0.35cm} $\mathcal{M}_{3}=\sigma_{2}\otimes\tau_{1}$ \hspace{0.35cm} & \hspace{0.25cm} spin triplet &\hspace{0.8cm} reflection symmetries broken &\hspace{0.55cm} $A_{2,\Gamma}^{+}$\\
 $\textrm{SC}_{4}$ & \hspace{0.35cm} $\mathcal{M}_{4}=\sigma_{2}\otimes\tau_{2}$ \hspace{0.35cm} & \hspace{0.25cm} spin singlet &\hspace{0.8cm} reflection symmetries broken &\hspace{0.55cm} $A_{2,\Gamma}^{-}$\\
 $\textrm{PWD}_{1}$ & \hspace{0.35cm} $\mathcal{M}_{5}=\sigma_{0}\otimes\tau_{1}$ \hspace{0.35cm} & \hspace{0.25cm} spin singlet &\hspace{0.8cm} pair-density-wave & \hspace{0.55cm} $A_{\pm K}^{+}$\\
 $\textrm{PWD}_{2}$ & \hspace{0.35cm} $\mathcal{M}_{6}=\sigma_{3}\otimes\tau_{2}$ \hspace{0.35cm} & \hspace{0.25cm} spin triplet  &\hspace{0.8cm} pair-density-wave & \hspace{0.55cm} $A_{\pm K}^{+}$\\
 $\textrm{PWD}_{3}$ & \hspace{0.35cm} $\mathcal{M}_{7}=\sigma_{3}\otimes\tau_{1}$ \hspace{0.35cm} & \hspace{0.25cm} spin singlet &\hspace{0.8cm}  pair-density-wave & \hspace{0.55cm} $A_{\pm K}^{-}$\\
 $\textrm{PWD}_{4}$ &\hspace{0.35cm}  $\mathcal{M}_{8}=\sigma_{0}\otimes\tau_{2}$ \hspace{0.35cm} & \hspace{0.25cm} spin triplet  &\hspace{0.8cm} pair-density-wave & \hspace{0.55cm} $A_{\pm K}^{-}$\\
\hline
\hline
\end{tabular}
}
\label{Table_II}
\end{table*}

\subsection{Fermion-bilinear source terms and susceptibilities}\label{Sec_V_A}

Based on the peculiar features of RTG systems, fermion-fermion interactions may drive a phase transition 
to another state~\cite{Vafek2014PRB2, Maiti2010PRB,Halboth2000RPL,
Nandkishore2012NP,Chubukov2016PRX,Roy2018PRX2}. As addressed in Ref.~\cite{Levitov2021CubicBand}, eight underlying
candidate states are associated with
this phase transition, as listed in Table~\ref{Table_II}.

\begin{figure}[htbp]
\centering
\includegraphics[width=3.0in]{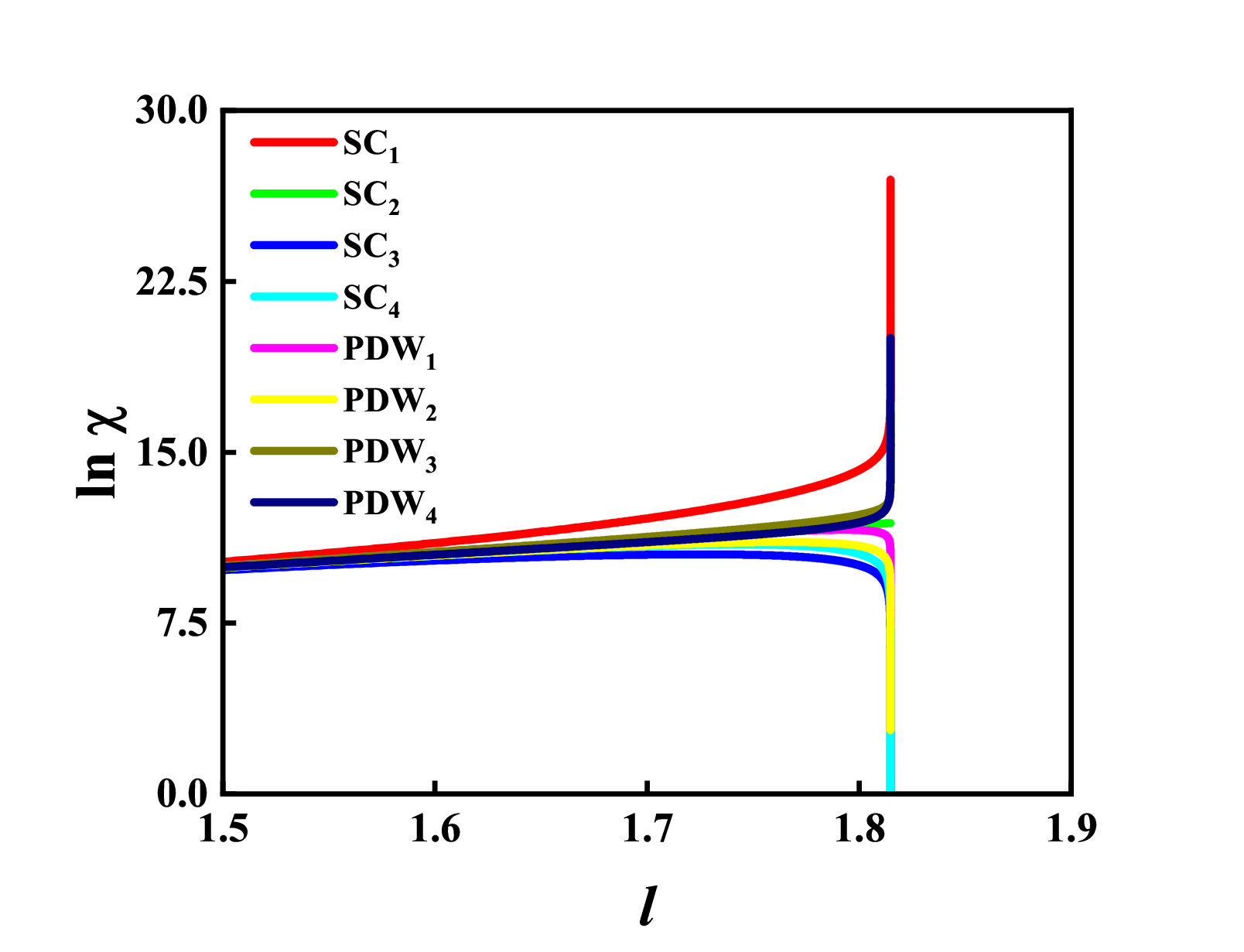}
\caption{(Color online) Energy-dependent susceptibilities induced by instabilities for all candidate states in Table~\ref{Table_II} around the $\textrm{FP}_{1}$ of the special case with a representative initial condition $D_{0} = 10^{-1}$
and $\lambda_{i0} = 10^{-1}$ with $i=1-6$ (the basic results are insusceptible to the concrete initial conditions).}\label{Fig_11}
\end{figure}

For the purpose of investigating these states, we introduce the following fermion-bilinear source terms
to represent the potential candidate states~\cite{Maiti2010PRB,Vafek2014PRB2,Chubukov2016PRX},
\begin{eqnarray}
S_{\textrm{sou}}
&=&\int d\tau\int d^{2}\mathbf{x}\left[\sum^8_{k=1}g_{k}\Psi^{\dag}\mathcal{M}_{k}\Psi^{\ast}+\mathrm{H.C.}\right],\label{Eq_5_1}
\end{eqnarray}
where the matrices $\mathcal{M}_{k}$ with $k=1-8$ are associated with the related candidate states distinguished by different symmetry breakings
as presented in Table~\ref{Table_II}. In addition, the parameters $g_{k}$ are employed to
measure the strengths of the corresponding source terms.

In order to examine the instabilities around the fixed points, we henceforth combine the
effective action~(\ref{Eq_S_eff}) and the source terms~(\ref{Eq_5_1}) to obtain a
renormalized effective action~\cite{Vafek2014PRB2, Maiti2010PRB,Halboth2000RPL,
Nandkishore2012NP,Chubukov2016PRX,Roy2018PRX2}
\begin{eqnarray}
S'_{\mathrm{eff}} = S_{\mathrm{eff}} + S_{\mathrm{sou}}.
\label{Eq_S_eff_new}
\end{eqnarray}
To proceed, we consider the renormalized effective action~(\ref{Eq_S_eff_new}) and then follow the procedures in Sec.~\ref{Sec_RG_eqations}
to derive the energy-dependent evolutions for the strengths of the source terms after taking into account the
one-loop corrections~\cite{Vafek2014PRB2,Chubukov2016PRX,Roy2018PRX2}, which can be formally expressed as
\begin{eqnarray}
\frac{dg_{k}}{dl}=\mathcal{J}_k(D,\lambda_i,g_{k}),\label{g_i_exp}
\end{eqnarray}
where $D,\lambda_i$ with $i=1-6$ appears in Eq.~(\ref{Eq_S_eff}), and Appendix~\ref{Appendix_source_terms}
provides the one-loop corrections and the concrete expressions of $\mathcal{J}_k$.
On the basis of these, we can derive the corresponding energy-dependent susceptibilities for these source terms, which are
designated by~\cite{Vafek2014PRB2}
\begin{eqnarray}
\delta\chi_k\equiv\frac{\partial^2 f}{\partial g_{k}(0)\partial g^*_k(0)},\label{Eq_chi}
\end{eqnarray}
where the variable $f$ serves as the free energy density, and $g_{k}$ corresponds to the strengths of source terms.

\begin{figure*}[htbp]
\centering
\includegraphics[width=2.5in]{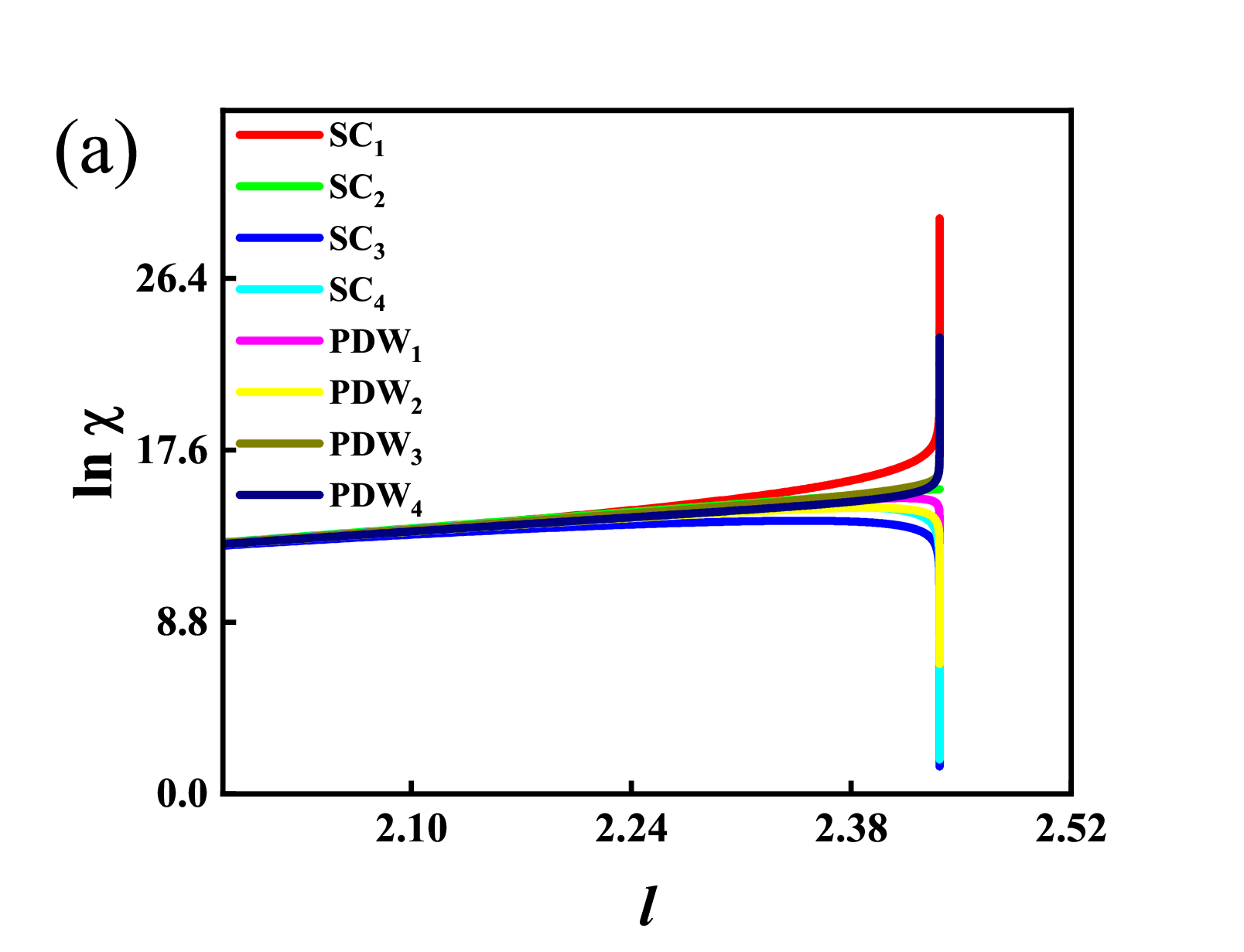}
\includegraphics[width=2.5in]{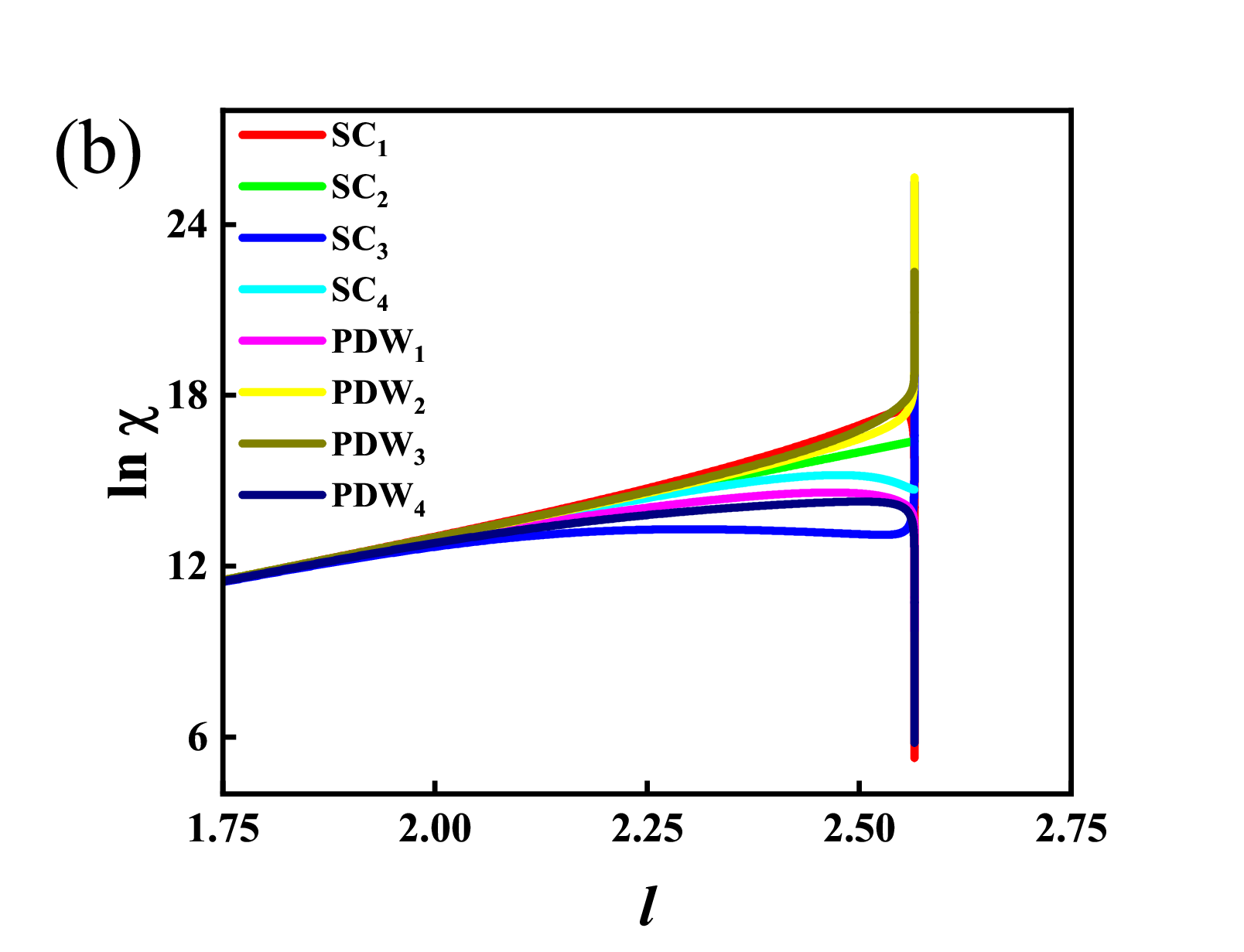}\\
\includegraphics[width=2.5in]{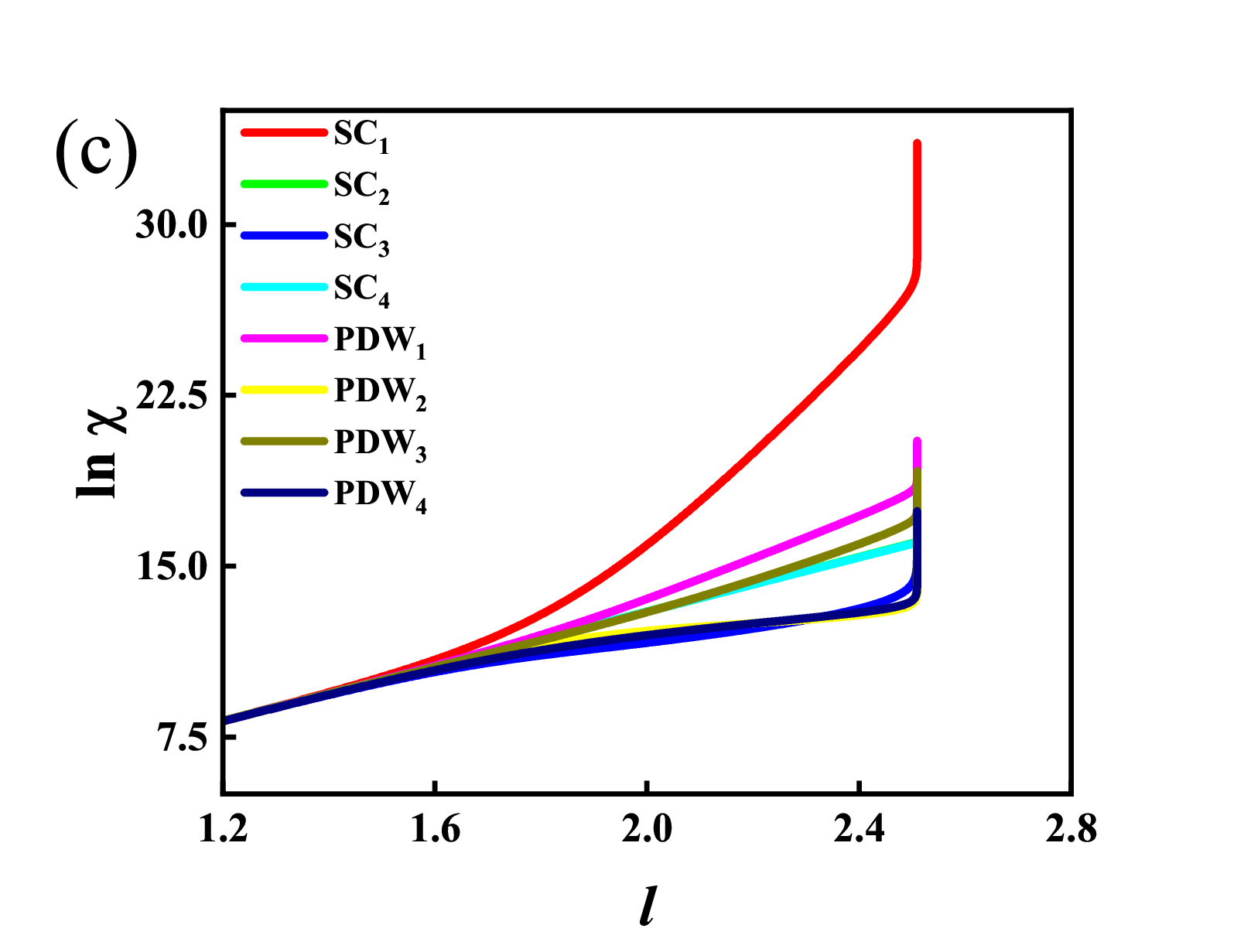}
\includegraphics[width=2.5in]{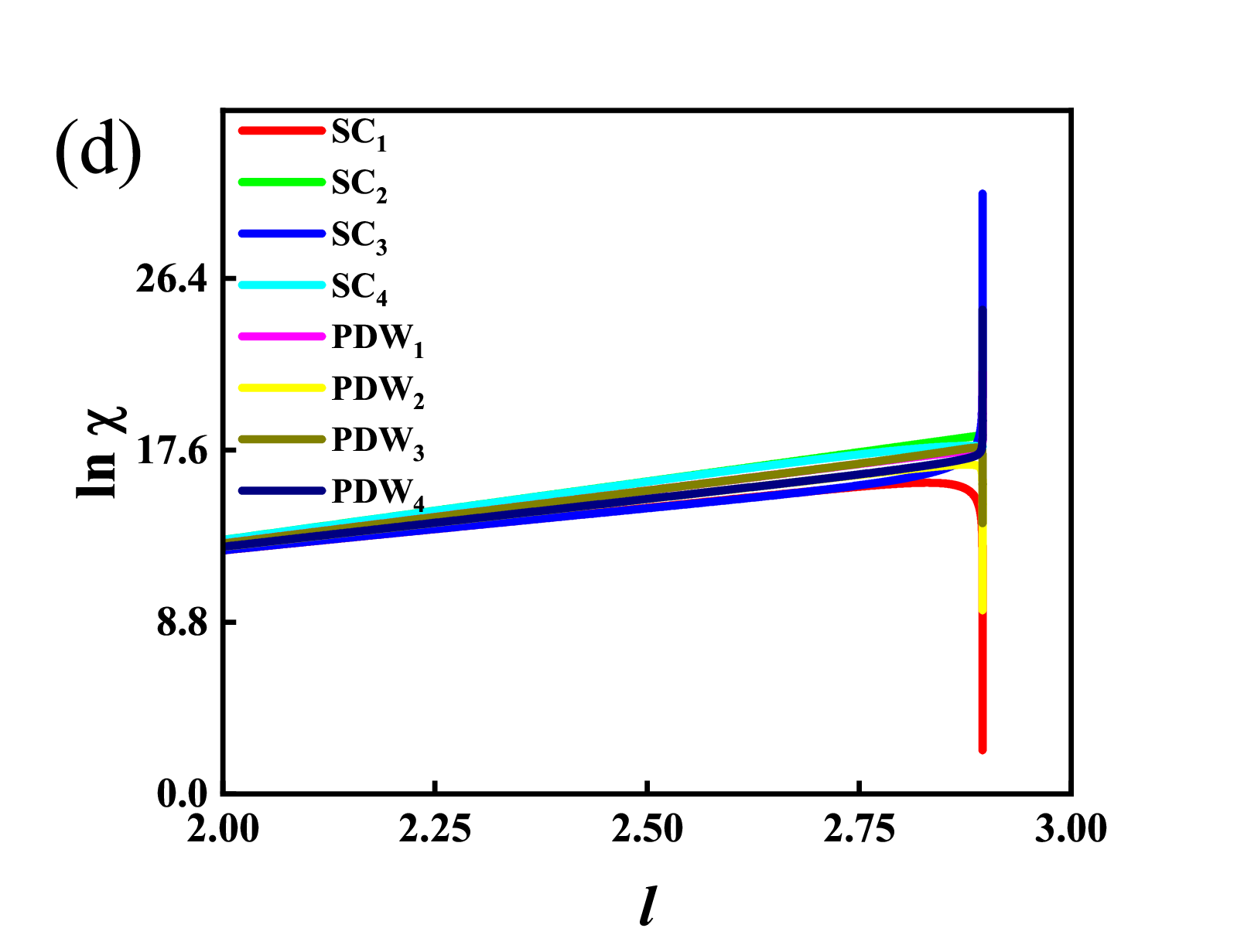}
\caption{(Color online) Energy-dependent susceptibilities induced by instabilities for all candidate states in Table~\ref{Table_II} around the fixed points of the general case: (a) $\textrm{FP}_{1}$, (b) $\textrm{FP}_{2}$,
(c) $\textrm{FP}_{3}$, and (d) $\textrm{FP}_{4}$, respectively.}\label{Fig_12}
\end{figure*}

\subsection{Favorable dominant states around the fixed points}\label{Sec_V_B}

Subsequently, let us consider the leading phases around the FPs.
As these FPs are approached, multiple competing candidate phases may exist. In principle,
susceptibility can serve as a key indicator for identifying the leading phases which are
associated with the most significantly divergent susceptibility during the phase transitions ~\cite{Halboth2000RPL,Maiti2010PRB,Vafek2014PRB2,Chubukov2016PRX}.
Consequently, we can calculate and compare the energy-dependent susceptibilities that are associated with each possible
symmetry breaking~\cite{Halboth2000RPL,Maiti2010PRB,Vafek2012PRB,Vafek2014PRB2,Chubukov2016PRX} to identify the most favorable states among all candidate states listed in Table~\ref{Table_II} as the distinct fixed points addressed in
Sec.~\ref{Sec_fixed_points} are approached. To this end, we need to consider the coupled RG equations of interaction parameters~(\ref{Eq_3_6})-(\ref{Eq_3_12}) and the energy-dependent flows of source strengths~(\ref{g_i_exp}) simultaneously.
Inserting these energy-dependent evolutions into the susceptibility formula~(\ref{Eq_chi}) then yields the energy-dependent susceptibilities.

First, let us consider the special case, in which there is only one fixed point, $\textrm{FP}_{1}$. As this fixed point is reached,
numerical analysis yields the energy-dependent susceptibilities in Fig.~\ref{Fig_11} with a representative set of initial values for $D_{0} = 10^{-1}$
and $\lambda_{i0} = 10^{-1}$ with $i=1-6$ (the basic results are insensitive to the specific initial conditions).
It can be clearly seen in Fig.~\ref{Fig_11} that the $\textrm{SC}_{1}$ state dominates over all the other candidates in Table.~\ref{Table_II} and
becomes the leading state for the phase transition induced by the instability around this fixed point.
In addition to the leading state, we observe that the $\textrm{PWD}_{4}$ is second only to $\textrm{SC}_{1}$ and can be regarded
as a subleading state--less significant than the leading state but strongly competing with other states. It may even become  dominant when the additional effects are considered.

Next, we shift our attention to the general case.
Under this circumstance, there exist four distinct types of fixed points exist as summarized in Table~\ref{Table_I}.
After performing a similar numerical analysis with several representative initial conditions,
we present the primary results in Fig.~\ref{Fig_12}.

As $\textrm{FP}_{1}$ is approached for the general case, which can be obtained by setting
$\lg\ \kappa < 2$, or $\lg\ \eta > -4$ or $\lg\ \zeta < -2$, Fig.~\ref{Fig_12}(a) indicates that the leading and the subleading states correspond to $\textrm{SC}_{1}$ and $\textrm{PDW}_{4}$, respectively, which are in good agreement with those in the special case.
Considering the initial conditions of the fourth and fifth rows in Table~\ref{Table_II}, the coupled RG equations~(\ref{Eq_3_6})-(\ref{Eq_3_12})
drive the RTG system to $\textrm{FP}_{3}$ and $\textrm{FP}_{4}$ at the lowest-energy limit, respectively.
By capturing the evolutions of all parameters during this process, Fig.~\ref{Fig_12}(c) and Fig.~\ref{Fig_12}(d)
show the associated energy-dependent susceptibilities of all the candidate states for these two fixed points. Clearly,
near $\textrm{FP}_{3}$, Fig.~\ref{Fig_12}(c) shows that $\textrm{SC}_{1}$ dominates over all other states and the $\textrm{PDW}_{1}$ as the subdominant state.
In comparison, $\textrm{SC}_{1}$ and $\textrm{PDW}_{1}$, are replaced by $\textrm{SC}_{3}$ and $\textrm{PDW}_{4}$,
as depicted in Fig.~\ref{Fig_12}(d) around $\textrm{FP}_{4}$. With respect to $\textrm{FP}_{2}$, we notice that
both the leading and subleading states are no longer superconducting states but instead PDW states. Specifically,
taking $D_{0}=10^{-1}, \lambda_{10}=10^{-7}, \lambda_{20}=10^{-2}, \lambda_{30}=10^{-7}, \lambda_{40}=10^{-5}, \lambda_{50}=10^{-2}$, and $\lambda_{60}=10^{-3}$ for an instance, which satisfies the $\textrm{FP}_{2}\mathrm{-Condition}$~(\ref{Eq_4_3}), Fig.~\ref{Fig_12}(b) exhibits the energy-dependent evolution of susceptibilities.  As a consequence, this clearly manifests that the leading and subleading phases are occupied by $\textrm{PDW}_{2}$ and $\textrm{PDW}_{3}$, respectively.

Further, it is interesting to address the competition between dominant and subdominant phases around the FPs.
It is convenient to measure this competition by designating the ratio $\gamma \equiv \frac{\chi_{\text{dom}}}{\chi_{\text{sub}}} \big|_{l_c}$, where $\chi_{\text{dom}}$ and $\chi_{\text{sub}}$ denote the susceptibilities of the dominant and subdominant phases
around the FPs, respectively.
This ratio clusters into two distinct cases as shown in Fig.~\ref{Fig_12}.
For Case-I with $\gamma < 1.5$, there is strong competition
between the dominant and subdominant phases. In this case, the dominant is not stable enough and may be superseded by the subdominant state under certain additional influences. In comparison, as for Case-II with $\gamma < 1.5$, the leading state exhibits clear dominance over the competing phase, leading to a relatively stable phase transition.
$\mathrm{FP}_1$, $\mathrm{FP}_2$, and $\mathrm{FP}_4$ belong to Case-I,
where strong competitive interactions between the subdominant and dominant phases.
These subdominant phases compete with the dominant phases and have the potential opportunities to become dominant
when additional perturbations are introduced to the related system.
Specifically, at $\mathrm{FP}_1$, both $\mathrm{PDW}_4$ and $\mathrm{PDW}_3$ are subordinate to
the $\mathrm{SC}_1$. Around $\mathrm{FP}_2$, the dominant phase shifts to $\mathrm{PDW}_2$,
while subdominant phases arise in $\mathrm{SC}_3$ and $\mathrm{PDW}_3$.
Near $\mathrm{FP}_4$, $\mathrm{SC}_3$ emerges as the dominant phase, with $\mathrm{PDW}_4$ and $\mathrm{PDW}_1$
as the subleading phases. In contrast, $\mathrm{FP}_3$ falls under Case-II,
where the dominant phase $\mathrm{SC}_1$ exhibits relatively higher stability compared to
all other candidate states.

To recapitulate, the fermion-fermion interactions play an important role in shaping the low-energy physics. In particular, their strong
competition can drive the RTG system to four distinct types of fixed points, as summarized in Table~\ref{Table_I}, depending on the
initial conditions. Furthermore, as the RTG system approaches these fixed points, certain instabilities emerge, leading to phase transitions into other states listed in Table~\ref{Table_II}. Specifically, the superconducting states dominate over PDW for the $\textrm{FP}_{1,3,4}$ and
vice versa for the $\textrm{FP}_{2}$.

\subsection{Comments and discussions}

Before closing this section, it is necessary to provide several comments and discussions.
To begin with, we present a brief comparison with previous studies.
Earlier theoretical investigations on multilayer graphene~\cite{Koshino2009PRB,Zhang2010PRB,Levitov2021CubicBand,You2022PRB,Cea2022PRB},
such as those in Ref.~\cite{Levitov2021CubicBand}, primarily focused primarily on local density-density interactions.
While significant results have been obtained from density-density interactions,
other types of fermion-fermion interactions are allowed from a theoretical standpoint.
Without these interactions, some key contributions may be neglected,
leading to an incomplete understanding of the low-energy behavior of the system.
By considering these types of interactions, we identify four distinct types of
fixed points and determine the dominant states around them, namely the $\mathrm{SC}_{1}$,
$\mathrm{SC}_{3}$, and $\mathrm{PDW}_{2}$ phases--providing a more comprehensive
understanding of the low-energy properties of RTG.
In particular, recent experimental studies~\cite{Zaliznyak2011NPhys,Zhang2010PRB,Zhou2021Nature,
MacDonald2203.12723,Weitz2024NPhys} indicate the presence of superconductivity in RTG,
which may correspond to the $\mathrm{SC}_{1}$ phase~\cite{Zhou2021Nature}, related to the dominant states
around $\mathrm{FP}_{1}$ and $\mathrm{FP}_{3}$. This motivates further study of related
quantities in similar materials in the future.

Next, we provide some comments on the issues related to RTG systems.
In multilayer graphene systems, several experiments suggested that the quality of nesting can be
tuned~\cite{Lu2022PRB,You2022PRB}.
Proper nesting can facilitate electron transitions between localized areas of the Fermi
surface and enhance the possibility to form the Cooper pairs via the Kohn-Luttinger mechanism~\cite{Lu2022PRB}.
This implies that there may be a link between
these nested ``hot spots" and the emergence of superconductivity.
In addition, the unique structure of RTG~\cite{Koshino2009PRB,Zhang2010PRB,Zhou2021Nature,Law2024arXiv,Levitov2024arXiv,Roy2109.04466,
MacDonald2203.12723,Weitz2024NPhys,Wilson2020PRR} introduces significant modifications to interlayer electronic interactions~\cite{Cea2022PRB,Xu2012PRB}, potentially leading to an antiferromagnetic (AFM) configuration.
Furthermore, the effects of external fields, such as the displacement fields, may play an important role in multilayer graphene systems by modifying interlayer distances and electronic structures~\cite{Qin2023PRL, Han2023JACS, Han2024JACS}. Specifically, displacement fields could induce energy differences between adjacent layers, significantly influencing the electronic band structure~\cite{Han2023JACS, Han2024JACS} and affecting physical behavior as well as topological properties in the low-energy regime~\cite{Qin2023PRL,Han2023JACS, Han2024JACS}.
In principle, these issues are highly intriguing yet inherently challenging. Addressing them in future studies would
help elucidate the mechanisms of critical behavior and improve our understandings of RTG systems and related materials.

\section{Summary}\label{Sec_summary}

In summary, the effects of fermion-fermion interactions on the low-energy physics of RTG materials are carefully investigated by
using the RG approach~\cite{Wilson1975RMP,Polchinski9210046,Shankar1994RMP}. Within the RG framework, we take into
account all the one-loop corrections and establish the coupled
RG equations for the fermionic couplings. Armed with these intertwined RG evolutions, detailed numerical analysis reveals
several interesting energy-dependent properties owing to the presence of distinct types of fermion-fermion interactions.

A comprehensive analysis indicates that the strong competition among different types of fermion-fermion
interactions is essential for determining the low-energy physics of the RTG system.
First, we obverse that the RTG system is attracted to four different types of fixed points in the interaction-parameter space--
$\textrm{FP}_{1}$, $\textrm{FP}_{2}$, $\textrm{FP}_{3}$, and $\textrm{FP}_{4}$--as discussed in Sec.~\ref{Sec_fixed_points}.
In particular, the initial conditions catalogued in Table~\ref{Table_I} are crucial in determining which fixed point that the system flows toward. Next, we find that these fixed points are typically associated with certain instabilities and  specific symmetry breaking, which can induce phase transitions to other states as summarized in Table~\ref{Table_II}.
After introducing the source terms and comparing the related  susceptibilities of all candidate states in Table~\ref{Table_II},
we find that the dominant phases near $\textrm{FP}_{1,2,3,4}$ correspond to $\textrm{SC}_{1}$, $\textrm{PDW}_{2}$, $\textrm{SC}_{1}$, and $\textrm{SC}_{3}$ states, respectively. These results serve as a necessary supplement and suitable extension to studies
that consider only density-density interactions~\cite{Levitov2021CubicBand}. We hope our findings provide
useful insights for further investigations into the low-energy properties of RTG and analogous materials.
Despite this progress, several intriguing but challenging questions remain unresolved, including
the mechanism of superconductivity in RTG, the interplay between antiferromagnetic and superconducting phases,
and the influence of displacement fields on phase transitions. Elucidating these issues in future would help uncover the
mechanisms of critical behavior and improve our understanding of the RTG systems and related materials.

\section*{ACKNOWLEDGEMENTS}

We thank Yi-Sheng Fu and Wen Liu for the helpful discussions.
J.W. is partially supported by the National Natural
Science Foundation of China under Grant No. 11504360.

\vspace{0.5cm}

\appendix

\section{One-loop corrections}\label{Appendix_1L-corrections}

According to the effective action~(\ref{Eq_S_eff}), the fermionic propagator does not receive nontrivial contributions from the
fermion-fermion interactions at the one-loop level~\cite{Vafek2014PRB2}. However, the fermion-fermion interactions themselves acquire  important
corrections, as shown in Fig.~\ref{Fig_1L-ff}. After lengthy but straightforward calculations, we obtain
the following one-loop corrections,
\begin{widetext}
\begin{figure}[htbp]
\centering
  \includegraphics[scale=1]{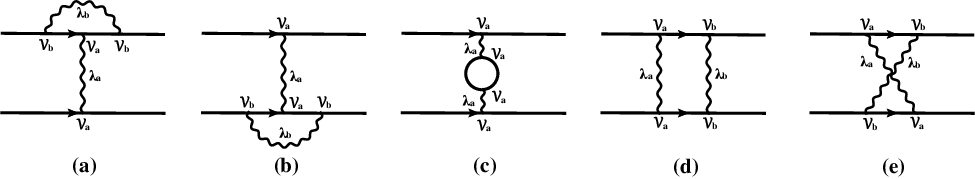}\\
  \caption{One-loop corrections to the fermion-fermion couplings (a)-(e) due to distinct types of combinations. The solid and wavy lines represent
  the free fermion propagator and fermion-fermion interaction, respectively.}\label{Fig_1L-ff}
\end{figure}
\begin{eqnarray}
\delta S_{\lambda_{1}}
&=&2\lambda_{1}[(\lambda_{1}-\lambda_{2}+\lambda_{3}+\lambda_{4}+\lambda_{5}+\lambda_{6})(\mathbb{A+B-C+D})
-4\lambda_{1}(\mathbb{A+B-C+D})]l\mathcal{P}_{1},\label{Eq_Appendix_1}\\
\delta S_{\lambda_{2}}
&=&\{-2\lambda_{2}[(\lambda_{1}-\lambda_{2}-\lambda_{3}+\lambda_{4}-\lambda_{5}+\lambda_{6})(\mathbb{A+B+C-D})-4\lambda_{2}(\mathbb{A+B+C-D})\nonumber\\
&&-2(2\lambda_{3}\lambda_{6}+2\lambda_{4}\lambda_{5})\mathbb{B}-2(\lambda_{1}^{2}+\lambda_{2}^{2}+\lambda_{3}^{2}
+\lambda_{4}^{2}+\lambda_{5}^{2}+\lambda_{6}^{2})\mathbb{D}-4\lambda_{1}\lambda_{2}\mathbb{C}\}l\mathcal{P}_{2},\\
\delta S_{\lambda_{3}}
&=&\{2\lambda_{3}[(\lambda_{1}+\lambda_{2}+\lambda_{3}+\lambda_{4}-\lambda_{5}-\lambda_{6})(\mathbb{A+B-C-D})-4\lambda_{3}(\mathbb{A+B-C-D})]\nonumber\\
&&-2(2\lambda_{2}\lambda_{6}\mathbb{B}+2\lambda_{2}\lambda_{3}\mathbb{D})\}l\mathcal{P}_{3},\\
\delta S_{\lambda_{4}}
&=&2\lambda_{4}[(\lambda_{1}-\lambda_{2}+\lambda_{3}+\lambda_{4}-\lambda_{5}-\lambda_{6})(\mathbb{A-B-C+D})-4\lambda_{4}(\mathbb{A-B-C+D})]l\mathcal{P}_{4},\\
\delta S_{\lambda_{5}}
&=&\{2\lambda_{5}[(\lambda_{1}+\lambda_{2}-\lambda_{3}-\lambda_{4}+\lambda_{5}+\lambda_{6})(\mathbb{A+B-C-D})-4\lambda_{5}(\mathbb{A+B-C-D})]\nonumber\\
&&-2(2\lambda_{2}\lambda_{4}\mathbb{B}+2\lambda_{2}\lambda_{5}\mathbb{D})\}l\mathcal{P}_{5},\\
\delta S_{\lambda_{6}}
&=&2\lambda_{6}[(\lambda_{1}-\lambda_{2}-\lambda_{3}-\lambda_{4}+\lambda_{5}+\lambda_{6})(\mathbb{A-B-C+D})
-4\lambda_{6}(\mathbb{A-B-C+D})]l\mathcal{P}_{6},\label{Eq_Appendix_2}
\end{eqnarray}
where the $\mathcal{P}_{k}$ with $k=1-6$ are defined as
\begin{eqnarray}
\mathcal{P}_{k}\!\equiv\!\int_{-\infty}^{\infty}\!\frac{d\omega_{1}d\omega_{2}d\omega_{3}}{(2\pi)^{3}}
\int\!\frac{d^{2}\mathbf{k}_{1}d^{2}\mathbf{k}_{2}d^{2}\mathbf{k}_{3}}{(2\pi)^{6}}
\Psi^{\dag}(\omega_{1},\mathbf{k}_{1})\mathcal{V}_{k}\Psi(\omega_{2},\mathbf{k}_{2})
\Psi^{\dag}(\omega_{3},\mathbf{k}_{3})\mathcal{V}_{k}\Psi(\omega_{1}+\omega_{2}-\omega_{3},\mathbf{k}_{1}+\mathbf{k}_{2}-\mathbf{k}_{3}),
\end{eqnarray}
and the related coefficients are denominated as
\begin{eqnarray}
\mathbb{A} & \equiv & \frac{1}{16\pi^{2}}\int_{0}^{2\pi}\left[-\frac{1}{\left(\cos^{2}{3\theta}+\beta^{2}\sin^{2}{3\theta}\right)^{1/2}}
+\frac{D^{2}}{2\left(\cos^{2}{3\theta}+\beta^{2}\sin^{2}{3\theta}\right)^{3/2}}\right]d\theta\\\label{Eq_A_1}
\mathbb{B} & \equiv & \frac{1}{16\pi^{2}}\int_{0}^{2\pi}\left[-\frac{\cos^{2}(3\theta)}{\left(\beta^{2}\sin^{2}(3\theta)+\cos^{2}(3\theta)\right)^{3/2}}
+\frac{3D^{2}\cos^{2}(3\theta)}{2\left(\beta^{2}\sin^{2}(3\theta)+\cos^{2}(3\theta)\right)^{5/2}}\right]d\theta\\\label{Eq_A_2}
\mathbb{C} & \equiv & \frac{1}{16\pi^{2}}\int_{0}^{2\pi}\left[-\frac{\sin^{2}(3\theta)\beta^{2}}{\left(\beta^{2}\sin^{2}(3\theta)+\cos^{2}(3\theta)\right)^{3/2}}
+\frac{3D^{2}\sin^{2}(3\theta)\beta^{2}}{2\left(\beta^{2}\sin^{2}(3\theta)+\cos^{2}(3\theta)\right)^{5/2}}\right]d\theta\\\label{Eq_A_3}
\mathbb{D} & \equiv & \frac{1}{16\pi^{2}}\int_{0}^{2\pi}\left[ -\frac{D^{2}}{\left(\beta^{2}\sin^{2}(3\theta)+\cos^{2}(3\theta)\right)^{3/2}}+\frac{3D^{4}}{2\left(\beta^{2}\sin^{2}(3\theta)
+\cos^{2}(3\theta)\right)^{5/2}}\right]d\theta\\\label{Eq_A_4}
\end{eqnarray}
\end{widetext}


\begin{figure}[htbp]
\centering
  \includegraphics[scale=0.5]{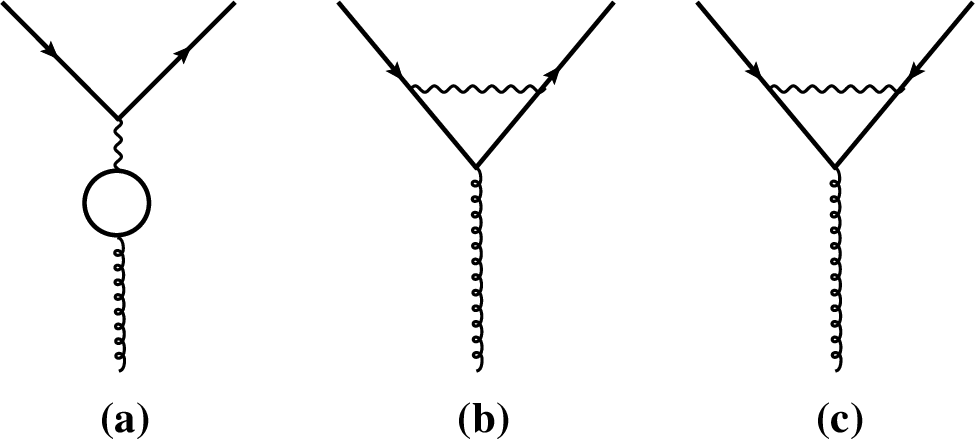}\\
  \caption{One-loop corrections to the fermion-bilinear
source terms. 
The solid, wave, and gluon lines
correspond to the fermionic, fermion-fermion interaction
and source term, respectively.}\label{Fig_1L_source}
\end{figure}

\section{One-loop flows of source terms}\label{Appendix_source_terms}

Starting with the renormalized effective action~(\ref{Eq_S_eff_new}) in the presence of source terms, one finds that three one-loop diagrams contribute to the strengths of source terms, as shown in Fig.~\ref{Fig_1L_source}. Carrying out
a similar analytical analysis utilized in Appendix~\ref{Appendix_1L-corrections} and performing the
RG analysis with the help of the RG rescalings~(\ref{Eq_scaling-1})-(\ref{Eq_scaling-4}) eventually yield the following flow equations
for the strengths of source terms with $k=1-8$:
\begin{eqnarray}
\frac{dg_1}{d l}
\!&=&\!2\left[3\!-\!\mathcal{T}_{1}
(\lambda_{2}\!-\!\lambda_{1}-\lambda_{3}-\lambda_{4}-\lambda_{5}-\lambda_{6})\right]g_1,\\\label{Eq_B_1}
\frac{dg_2}{d l}
\!&=&\!6g_2,\\\label{Eq_B_2}
\frac{dg_3}{d l}
\!&=&\!2\left[3
\!-\!\mathcal{T}_{1}(\lambda_{2}\!-\!\lambda_{1}+\lambda_{3}+\lambda_{4}+\lambda_{5}+\lambda_{6})\right]g_3,\\\label{Eq_B_3} 
\frac{dg_4}{d l}
\!&=&\!6 g_4,\\\label{Eq_B_4}
\frac{dg_5}{d l}
\!&=&\!2\left[3
\!-\!\mathcal{T}_{2}(\lambda_{2}\!-\!\lambda_{1}-\lambda_{3}-\lambda_{4}+\lambda_{5}+\lambda_{6})\right]g_5,\\\label{Eq_B_5}
\frac{dg_6}{d l}
\!&=&\!2\left[3
\!-\!\mathcal{T}_{3}(\lambda_{1}\!-\!\lambda_{2}+\lambda_{3}-\lambda_{4}-\lambda_{5}+\lambda_{6})\right]g_6,\\\label{Eq_B_6}
\frac{dg_7}{d l}
\!&=&\!2\left[3
\!-\!\mathcal{T}_{2}(\lambda_{2}\!-\!\lambda_{1}+\lambda_{3}+\lambda_{4}-\lambda_{5}-\lambda_{6})\right]g_7,\\\label{Eq_B_7}
\frac{dg_8}{d l}
\!&=&\!2\left[3
\!-\!\mathcal{T}_{3}(\lambda_{1}\!-\!\lambda_{2}-\lambda_{3}+\lambda_{4}+\lambda_{5}-\lambda_{6})\right]g_8,\label{Eq_B_8}
\end{eqnarray}
where the coefficients $\mathcal{T}_{1,2,3}$ and $\mathcal{M}$ as well as $\mathcal{N}$ are defined as
\begin{eqnarray}
\mathcal{T}_{1}&\equiv&\int_{0}^{2\pi}\frac{(3D^{2}-2\mathcal{M})}{32\pi^{2}\mathcal{M}^{\frac{3}{2}}}d\theta,\\\label{Eq_B_9}
\mathcal{T}_{2}&\equiv&\int_{0}^{2\pi}\frac{(3D^{2}-2\mathcal{M})
(\mathcal{M}-\mathcal{N})}{64\pi^{2}\mathcal{M}^{\frac{5}{2}}}d\theta,\\\label{Eq_B_10}
\mathcal{T}_{3}&\equiv&\int_{0}^{2\pi}
\frac{(3D^{2}-2\mathcal{M})(\mathcal{M}+\mathcal{N})}{64\pi^{2}\mathcal{M}^{\frac{5}{2}}}
d\theta,\\\label{Eq_B_11}
\mathcal{M}&\equiv&(\cos{\theta}^{3} -3\cos{\theta}\sin{\theta}^{2})^{2}\nonumber\\
&& +\beta^{2}(3\cos{\theta}^{2}\sin{\theta}-\sin{\theta}^{3})^{2},\\\label{Eq_B_12}
\mathcal{N}&\equiv&(\cos{\theta}^{3} -3\cos{\theta}\sin{\theta}^{2})^{2} \nonumber\\
&&-\beta^{2}(3\cos{\theta}^{2}\sin{\theta}-\sin{\theta}^{3})^{2}.\label{Eq_B_13}
\end{eqnarray}



\end{document}